\documentstyle[12pt]{article}
\begin{document}
\newcommand{\mycomment}[1]{\footnote{{\bf\sl #1}}}
\newcommand{\qq}{{\sl {\bf [\&:-)] }}\,}
\newcommand{\wg}{\wedge}
\newcommand{\expb}[1]{\exp\left\{ #1 \right\} }
\def\cN{{\cal N}}
\def\cG{{\cal G}}
\def\cK{{\cal K}}
\def\cL{{\cal L}}
\def\cO{{\cal O}}
\def\cW{{\cal W}}
\def\cC{{\cal C}}
\def\cT{{\cal T}}
\def\cZ{{\cal Z}}
\def\cR{{\cal R}}
\def\cA{{\cal A}}
\def\dd{{\rm d}}
\newcommand{\LL}{{I\!\! L}}
\newcommand{\be}{\begin{eqnarray}}
\newcommand{\ee}{\end{eqnarray}}
\newcommand{\Z}{{Z \!\!\! Z}}
\newcommand{\half}{\frac{1}{2}}
\newcommand{\vx}{{\bf x}}
\newcommand{\vy}{{\bf y}}
\newcommand{\eq}[1]{(\ref{#1})}
\newcommand{\diff}{\partial}
\newcommand{\beq}{\begin{equation}}
\newcommand{\eeq}{\end{equation}}
\newcommand{\beqn}{\begin{eqnarray}}
\newcommand{\eeqn}{\end{eqnarray}}
\newcommand{\ddd}{{\rm d}}
\newcommand{\cD}{{\cal D}}
\newcommand{\dddd}{\cD}
\newcommand\Appendix[1]{\par
 \setcounter{section}{0}
 \setcounter{equation}{0}
 \renewcommand{\thesection}{Appendix \Alph{section}}
 \section{#1}
 \def\theequation{\Alph{section}.\arabic{equation}}}
\def\cZ{{\cal Z}}
\def\cS{{\cal S}}
\def\cL{{I\!\! L}}
\def\dd{{\rm d}}  
\def\NP{ Nucl.~Phys.}
\def\PL{ Phys.~Lett.}
\def\PRL{ Phys.~Rev.~Lett.}



\vspace{10mm}


\centerline{\Large\bf A Correspondence Between Supersymmetric}
\centerline{\Large\bf Yang-Mills and Supergravity Theories}

\vspace{5mm}

\centerline{E.T.~Akhmedov\footnote{ Tel.: (7-095)129-9768.
Fax (7-095) 129-9649 \\ E-mail: akhmedov@itep.ru; akhmedov@physics.ubc.ca}}

\vspace{5mm}

\centerline{Institute of Theoretical and Experimental Physics}

\centerline{Moscow, 117259, B. Cheremushkinskaya, 25.}

\vspace{5mm}

\centerline{and}

\vspace{5mm}

\centerline{University of British Columbia}

\centerline{6224 Agricultural Rd, Vancouver BC, Canada, V6T 1Z1}

\vspace{10mm}

\begin{abstract}
AdS/CFT-correspondence establishes a
relationship between supersymmetric gravity (SUGRA) on Anti-de-Sitter
(AdS) space and supersymmetric Yang-Mills (SYM) theory
which is conformaly invariant (CFT). The AdS space is the solution
of the Einstein-Hilbert equations with a constant negative curvature.
Why is this relationship important? What kind of relationship is this?
How does one find it? The purpose of this text is to answer these
questions. We try to present the
main ideas and arguments underlying this relationship,
starting with a brief sketch
of old string theory statements and proceeding with the definition of
D-branes and a description of their main features.  We finish with the
observation of the correspondence in question and the arguments that favor
it.  \end{abstract}

\newpage

\tableofcontents

\newpage

\section{Introduction}

A long standing problem is to find a string description of
Yang-Mills (YM) theory \cite{Polyakov}.  The argument goes as follows:
In the mathematical description of any phenomenon one needs to find some exactly solvable approximation to it
and some small parameter which one can expand over in order to approach the
real situation.  In the case of YM theory such a good approximation at large
energies is in terms of free vector particles.  They carry quantum numbers
taking values in the adjoint representation of a non-Abelian gauge group and
the small parameter in question is the coupling constant $g^2$
\cite{Polyakov}. However, there is a problem in this description because quantum effects
cause the coupling constant to grow as one approaches large distances or small
energies.  Furthermore, at some distance scale the description in terms of the
fundamental YM variables becomes invalid due to singularities in the perturbative
theory \cite{Polyakov}.  As a result, nobody knows how to pass to low
energies in YM theory.  Thus the question appears: What is the
approximation to YM theory which can be applicable in both regimes
(high and low energies)?

  The most promising approach to this problem is to consider SU$(N)$ YM
theory as $N \to \infty$ \cite{Polyakov}.  In this limit the YM perturbation
series drastically simplifies \cite{tHooft} and the only graphs which survive
look like ``triangulations'' of sphere.  This is one of the hints
\cite{Polyakov} that there could be a string description of
YM theory in this limit, which would be a two-dimensional
theory representing these "triangulations".  Furthermore graphs which
contribute to the "triangulations" represent a series expansion in powers of
$g^2 N$ (which is taken finite as $N\to\infty$) rather than simply in
powers of the YM coupling constant $g^2$ \cite{tHooft}, while at the same time the torus
and all graphs having topologies of spheres with more than one handle are
suppressed by powers of $1/N^2$.  Thus $1/N^2$ would be the small parameter
in expansion over which one could approach the real situation.

   Why should we prefer the description in terms of string theory?  The
point is that string theory has a very well developed and powerful apparatus
for calculating amplitudes of various processes
\cite{Polyakov,GrScWi,Polchinski} and, moreover, one could hope to solve it.

   In the case of ordinary YM no one has yet succeeded in finding such a
string description, but in conformaly invariant supersymmetric YM (SYM)
theories there has recently been a considerable progress
\cite{Pol97,Maldacena,GuKlPo,Witten}. It is worth mentioning that
the string description of conformal YM theories is of pure academic
interest since due to conformal invariance we know dynamics of these theories at all
distances, however one might hope that such a string description would reveal some features of
string theory for the ordinary YM.

 There are several non-anomalous and self-consistent string theories which
satisfy supersymmetry (SUSY) in target space --- the space where the string
evolves.  The target space should be ten-dimensional, since otherwise
there is no well developed apparatus of calculation of superstring
amplitudes \cite{Polyakov,GrScWi,Polchinski}.  At the same time, string
world-sheets are two-dimensional universes swept out by strings during their time
evolution.  There are infinitely many ways to excite the world-sheet theory
to give different quantum states of the string.  Each of them looks like a
particle living in the target space. Among these particles there are a finite
number of massless ones, while all other particles have masses of the order
of the string tension, which is usually taken to be very big.
Hence, at distances bigger than the string characteristic one (exactly when the strings look as
point-like objects) only the massless particles survive, and the latter are
described by a field theory in the target space rather than string theory.

  Among the massless {\it closed} string excitations there is a symmetric
tensor particle, which, due to the symmetry properties of string theory, has
exactly the same number of degrees of freedom as the graviton.  The only
large scale theory (that which contains the lowest powers of
derivatives of the fields) which could describe the
graviton is Einstein-Hilbert gravity in the target space.  As can be
rigorously shown \cite{Polyakov,GrScWi,Polchinski} it is this theory
(interacting with the other massless string excitations) that follows from
string theory at large distances.  At the same time in the case of superstring
theory one obtains SUGRA at large distances.

   It is also possible to obtain SYM interacting with SUGRA if one
includes open strings in the theory along with closed ones.  This is because
the massless excitation of  {\it open} string theory is a vector particle
with the proper number of physical polarizations to be a gauge boson.

   Bearing this in mind, one could say that there is a string
theory for four-dimensional SYM. In the situation under consideration,
superstring theory gives a regularization of SYM theory. In fact, superstring
theory is finite and valid at any distances, while at large distances
it leads to a theory containing SYM. But this is
unsatisfactory because at the string characteristic scale when we get such a superstring
description of SYM we also have to deal with quantum gravity
and the dimensionality of space-time is ten rather than four.

   Fortunately new ways have recently been discovered to add open
string sectors to the closed ones.  They lead to new ways of coupling SYM
to SUGRA. To find them one has to add a stack of $N$
D-branes to closed string theory {\it in such a way that it respects SUSY}.
The D-branes are multi-dimensional sub-manifolds of the
target space on which open strings terminate \cite{PolD}, while closed strings can
still live in the bulk of target space.  Hence, in the world-volume of
a stack of $N$ D-branes at low energies one sees U$(N)$ SYM theory \cite{Wit95},
while in the bulk there is standard SUGRA.

  Thus, strings which could describe SYM theory are attached to our
four-dimensional world (D3-brane world-volume) while fluctuating in the bulk
of target space \cite{PolD,Pol97}.  To specify the string description one has
to find the geometry in which the strings fluctuate, and to do this one must
probe the D3-brane from the outside: from the bulk.  At the same time, the
theory as seen by an observer when traveling further and further from the
D-brane could change uncontrollably, for we do not know the full dynamics of
string theory.  To overcome this difficulty one has to force the
D-branes to respect some part of the SUSY transformations of superstring
theory. Let us explain why.

  In quantum field theory and statistical mechanics when one goes from small to
large distances, it is necessary to average over all fluctuations in the theory
with wave-lengths smaller than the distance scale in question.  This could lead to a
change of parameters in the theory.  For example, if one places a charged
source into a plasma it is screened, because opposite charges to the source
are attracted, while the same are repelled. It
is the simplicity of the system which allows us to predict how the charge of
a source will vary as one approaches it.  However, in YM theory the situation
is more complicated. In fact, how the charge of the theory
varies with respect to the distance is known only up to some low energy
scale.  As a result, a proper low energy description of strong interactions
is still unknown.  Similar things could happen in any non-linear theory,
such as gravity or string theory.

  The difference in the presence of SUSY is that bosonic and fermionic
degrees of freedom can be exchanged in the theory \cite{WeBa}.  It is this
symmetry which causes the cancellation between screening and anti-screening
due to fermions and bosons.  The latter happens only if a source respects
some part of SUSY transformations, i.e. that it is a
Bogomolni-Prassad-Sommerfeld (BPS) state \cite{Polchinski,WiOl,WeBa}.
While not rigorous, we hope that the reader has at least gotten the flavor of how it works.

  Thus, the presence of SUSY helps one to find how geometry is curved in the
D-brane background.  For example, the characteristic curvature of the
D3-brane is proportional to $\left(g^2 N\right)^{1/4}$ in the
units of string tension.

  Let us now explain the new ideas D-branes can give in seeking for string
description of YM theory in contrast to ``old'' string theory. First, in this case one could deal directly with
four-dimensional SYM theory. Second, one can vary the
regularization energy scale for YM theory living on a D-brane
world-volume and make it much smaller than the string one \cite{Pol97}.  This
works as follows:  After a regularization we have suppressed the
information about high frequency modes and the high energy theory underlying
the one in question should contain this information.  What happens in the
D-brane case is that high enough frequency modes of the fields living on
the D-brane world-volume could escape to the bulk of the target space:  They could
create closed strings living in the bulk \cite{Klebanov}.  However,
closed strings with energies smaller than the brane curvature can not
escape to infinity \cite{MaSt} but instead they stay in the throat
region --- the strongly curved part of the bulk in the vicinity of
the D-brane, because they do not have enough energy
to climb over the gravitational potential and escape to
the flat asymptotic region.

 In conclusion, we only need the theory in the throat in order to respect
unitarity \cite{Pol97,Maldacena,GuKl}.  Thus, if the limit $N\to\infty$ and
$g \to 0$ is taken in such a way that $g^2 N \ll 1$ we have the full
string theory in the throat regularizing the SYM on the D-brane world-volume
\cite{Maldacena} because this is the limit when the size of the D-brane
throat is very small and only string theory could be applicable.  However,
if limits $N\to\infty$ and $g \to 0$ are taken so that
$g^2 N \gg 1$ we could apply classical
gravity.  In fact, the size of the throat is very big in this situation.  We
mean that in this limit the string theory for SYM is described by the
{\it classical} superstring in the throat background, i.e. there is the
string description of SYM before gravity becomes quantized. Now in the
simplest situation SYM on the brane has $\cN = 4$ supersymmetries,
hence its $\beta$-function is zero and it is conformaly invariant.
In the corresponding gravity description the geometry of the throat of the
brane in question is AdS.

  The paper is organized as follows:  For self-consistency we include two chapters
devoted to string theory.  In the first chapter we try to present the main
ideas of string theory using the example of bosonic string theory, then in
the second chapter we proceed with the definition of type {\rm II} superstring
theories and review their massless spectrum. After presenting
superstring theory, we introduce the notion of D-branes in the third chapter
and show their relation to gravity solitons and to SYM theory.  We conclude
with the AdS/CFT-correspondence.  For completeness we have included the
discussion of the BPS states in an appendix.

   Unfortunately, it is impossible to present in detail the whole of
these subjects even in a big book, hence, our presentation is sketchy. We
hope that it highlights the main ideas and gives some food for thought about
the matters in question.  We are not trying to review this broad subject, and
our reference list is therefore far from complete;  A more or less complete
one can be found in \cite{RMAdS}.

\section{Bosonic string theory}

At present only first quantized string theory
\cite{Polyakov,GrScWi,Polchinski} is fully constructed.
This is ``quantum mechanics'' of string world-sheets, which are two-dimensional
spaces swept by quantum strings during their time evolution inside target space.
As for the relativistic particle the action for the
relativistic string is proportional to the area of its world-sheet:

\beqn
S \propto \int {\rm d}^2\sigma \, \sqrt{-\det|g_{ab}|},
\quad g_{ab} = \diff_a \tilde{x}_{\mu} \, \diff_b \tilde{x}_{\mu}, \label{ngstr}
\eeqn
where $\sigma_a$ $(a=1,2)$ are coordinates on the world-sheet
and $\tilde{x}_{\mu}(\sigma)$ $ (\mu = 0,d-1)$ are two-dimensional
functions describing the embeddings of strings into $d$-dimensional flat
target space.

However, the action \eq{ngstr} is nonlinear and, hence, difficult to quantize.
To make it quadratic in $\tilde{x}_{\mu}$ one includes a new dynamical
variable into the theory --- the string intrinsic metric $h_{ab}$ \cite{Polyakov}.
In this case string theory is described by a two-dimensional $\sigma$-model
interacting with two-dimensional gravity:

\beqn
S_{\rm st} = \frac{1}{2\pi\alpha'} \, \int {\rm d}^2\sigma \, \sqrt{-h}
\, h^{ab} \, \diff_a \tilde{x}_{\mu} \, \diff_b \tilde{x}_{\mu} +
\frac{1}{2\pi\alpha'} \, \int {\rm d}^2\sigma \, \sqrt{-h}, \nonumber \\
h = \det|h_{ab}|, \quad h^{ab} \propto h^{-1}_{ab}. \label{strac}
\eeqn
Here $\alpha'$ is the inverse string tension. Usually it is taken to be much
smaller than any distance scale which has been probed by experiments so far.

  On the level of classical equations of motion $h_{ab} \propto g_{ab}$.
Hence, the action \eq{strac} is classically equivalent to that in eq.
\eq{ngstr}.  On the quantum level, however, these two theories at least naively are
different (see, however, \cite{Polyakov}). In fact, in the functional integral of
theory \eq{ngstr} there is a summation over all possible string
world sheets, i.e. over embeddings $\tilde{x}_{\mu}$.  While in theory \eq{strac}
the sum is taken over all possible metrics on each world-sheet and over the
world-sheets themselves.

From now on we will be dealing with theory \eq{strac}. This theory is
invariant under the reparametrization transformations:
$\sigma_a \to f_a(\sigma)$, which represent general covariance on the
string world-sheets. Via this {\it two-parametric} symmetry one could get rid of
{\it two} components of the metric:

$$
{\rm d}s^2 = h^{ab} \, {\rm d}\sigma_a \, {\rm d}\sigma_b =
\exp{\left[\varphi(z,\bar{z})\right]} \, {\rm d}z \, {\rm d}\bar{z},
\quad {\rm where} \quad z = \exp{\left[\sigma_1 \, +
\, {\rm i} \, \sigma_2\right]}.
$$
For a world-sheet
with spherical topology this can be done unambiguously, while for the torus
and higher topologies, this can be done only up to a complex structure
\cite{GrScWi,Polchinski}. We are not explaining the details of the complex
structure, because we are not going to use this notion (except the fact that
it exists) anywhere below.

After the above reparametrization gauge fixing our action
is still invariant under the conformal transformations:

$$
z \to f(z), \quad \exp{\left[\varphi(z,\bar{z})\right]} \to
|\diff_z f(z)|^2 \, \exp{\left[\varphi(z,\bar{z})\right]}, \quad
\diff_{\bar{z}} \, f(z) = 0
$$
Using them it is possible to get rid of the internal metric giving

\beqn
S'_{\rm st} = \frac{1}{2\pi\alpha'} \, \int {\rm d}^2 z \,
\diff_{z} \tilde{x}_{\mu} \, \diff_{\bar{z}} \tilde{x}_{\mu} + \, {\rm
Faddeev-Popov \,\, ghost \,\, terms}. \label{strac1}
\eeqn
We do not discuss here what are Faddeev-Popov ghosts because we do not use
this notion (except the fact that it exists) anywhere below. A more or less
compleete discussion of the Faddeev-Popov ghosts within the string theory
one can find in \cite{GrScWi}.

   The possibility of getting rid of the metric as in \eq{strac}
is completely true only classically. In fact, after quantization of
the $\sigma$-model \eq{strac} there is the so called conformal anomaly
\cite{Polyakov,GrScWi,Polchinski}, because conformal symmetry is broken
by quantum effects. Hence, $\varphi(z,\bar{z})$ becomes a
dynamical field. It is necessary to cancel the anomaly since otherwise it is not
known how to calculate string theory correlation functions
\cite{Polyakov}. Contributions to the anomaly coming from
$\tilde{x}_{\mu} \,\,\, (\mu=0,...,d-1)$ and from the Faddeev-Popov ghosts
cancel each other if $d=26$.

   Furthermore there is a remnant of reparametrization invariance on string
world-sheets with higher topologies, which is referred to as modular
invariance. The modular transformations act on the complex structures
\cite{GrScWi,Polchinski}. One must also respect this invariance, because
otherwise there could be problems with gravitational and gauge anomalies in
the target space, and hence unitarity would be violated.

\subsection{Generating Functional}

   Only if all these symmetries are respected can one properly define
the fundamental quantity of bosonic string theory:

\beq
Z\left(\phantom{\frac12} G,B,{\it \Phi},T \,\, \right) = \sum^{\infty}_{\cG=0} Z_{\cG} = \sum^{\infty}_{\cG=0}
\int [\cD h_{ab}]_{\cG} \cD \tilde{x}_{\mu}
\exp{\left[- {\rm i} \,
S_{\rm st}\left(\tilde{x}_{\mu}^{\phantom{\frac12^{\frac12}}},
h_{ab}, G_{\mu\nu}, B_{\mu\nu}, {\it \Phi}, T\right)\right]}. \label{str}
\eeq
Here the measure $\prod_{\mu = 0}^{d-1} \, \cD\tilde{x}_{\mu}$ is as for $d$ scalars, while $[\cD
h_{ab}]_{\cG}$ should be defined in accordance with reparametrization, conformal and
modular invariances \cite{Polyakov}.

   The sum in this formula is over the genus $\cG$ of the string world-sheets.
This is an expansion over string loop corrections which are present in
addition to the aforementioned $\sigma$-model quantum corrections.
If one considers only closed strings these correction are represented by
spheres with $\cG$ handles, otherwise they are discs with holes and handles of
the total number\footnote{It is worth mentioning at this point that open
string theory contains closed one on its loop level. In fact, the annulus
amplitude (the first loop correction in open string theory) is
equivalent to the cylinder amplitude (the tree level in
closed string theory). Beside that, unitarity demands that closed string excitations should be
added to the ones of open string.} $\cG$.

  We start with a discussion of closed bosonic string theory.
In this case the action in \eq{str} is:

\beqn
S_{\rm st} \left(\tilde{x}_{\mu}^{\phantom{\frac12}}, h_{ab}, G_{\mu\nu}, B_{\mu\nu},
{\it \Phi}, T\right)
= \frac{1}{2\pi\alpha'}\int
{\rm d}^2\sigma \, \left\{\sqrt{-h} \, h^{ab} \, G_{\mu\nu}(\tilde{x}) \, \diff_a
\tilde{x}^{\mu} \, \diff_b \tilde{x}^{\nu} + \phantom{\frac12^{\frac12}} \right.
\nonumber\\ \left. \phantom{\frac12^{\frac12}} + \epsilon^{ab} \,
B_{\mu\nu}(\tilde{x}) \, \diff_a \tilde{x}^{\mu} \, \diff_b \tilde{x}^{\nu} +
\alpha' \, \sqrt{-h} \, R^{(2)} \, {\it \Phi}(\tilde{x}) + \sqrt{-h} \, T(\tilde{x})\right\},
\label{ac}
\eeqn
where $\epsilon^{ab}$ is the completely anti-symmetric tensor in two
dimensions and $R^{(2)}$ is two-dimensional scalar curvature for the
metric tensor $h_{ab}$.

   Now we see that dilaton's VEV ${\it \Phi}_\infty$
gives a coupling constant for the string loop expansion:

\beq
Z_{\cG} \propto \exp{\left\{- {\rm i}\frac{{\it \Phi}_{\infty}}{2\pi}\int
{\rm d}^2\sigma\sqrt{-h}
R^{(2)}\right\}} = \exp{\left[2 \, (\cG - 1) \,
{\it \Phi}_\infty\right]} = g_{\rm s}^{2 \, (\cG - 1)},
\eeq
Furthermore substituting $G_{\mu\nu} = \eta_{\mu\nu}$, $B_{\mu\nu} = 0$,
${\it \Phi} = 0$ and $T = 1$ into \eq{ac}, one gets for $S_{\rm st}$ the former
expression \eq{strac}.

 The physical meaning of $Z\left(G,B,{\it \Phi},T\right)$ is that it is the
generating functional for amplitudes of interactions between the {\it
smallest mass} string states.  In fact, we can get such amplitudes via
variations of the functional $Z$ over the sources $G,B,{\it \Phi}$ and $T$.
We are interested only in the smallest mass states because we
need to find a classical limit (large distance behavior) of string theory.
It is exactly this limit where we would use what is known from our world.  This is the reason
why we do not include any other sources, which would correspond to massive states,
into the functional integral
\eq{str}, \eq{ac}.

\subsection{Low energy spectrum}

    Why do the operators in \eq{str}, \eq{ac} with the sources
$G,B,{\it \Phi}$ and $T$
correspond to the smallest mass states? First it is necessary to explain how a
two-dimensional operator is related to a string state. The action of
an operator on the vacuum of the conformal theory \eq{strac} excites it. If
we take a particular harmonic of a source (for example,
$T = \, :\exp{\left[ {\rm i} \,
p_{\mu} \, \tilde{x}_{\mu}\right]}:$), it looks, from the target space point of view, as
some moving string in a particular quantum state. In fact, it is a plane wave
inside target space.

  Furthermore it is not necessary to modify the functional integral \eq{str}
to describe interactions of string states. This is one of the main
difference between string theory and a field theory describing particles. It relies on two
fundamental facts: First, in contrast to particle paths, for any disconnected
set of one-dimensional manifolds it is always possible to find a
two-dimensional string world-sheet which has these manifolds as components
of its boundary.  Such a world-sheet represents a Feynman graph for a string
amplitude and components of its boundary represents initial and final states of a process
in string theory.  Second, because of conformal symmetry, one can always
amputate external legs in a string amplitude. This is to say that by a
conformal transformation it is possible to shrink them, and each component of
the one-dimensional boundary to points on the world-sheet where
corresponding vertex operators are acting.

   To show why the operators in question correspond to the smallest mass
excitations, let us consider an $N$-point correlation function
\cite{Polyakov}:

\beqn
\cA_N = \int \prod^N_{j=1}
{\rm d}^2\sigma_j \, \left\langle \cO_1\Bigl(\tilde{x}_{\mu}(\sigma_1)\Bigr) \,
... \, \cO_N\Bigl(\tilde{x}_{\mu}(\sigma_N)\Bigr) \right\rangle,\label{massp}
\eeqn
where the average $\langle ... \rangle$ is taken with the functional
integral \eq{str}, where the action is as in \eq{strac}. Also, $\cO_j$ are some operators
with conformal weights\footnote{The definition of the conformal weight
$\Delta_j$ of an operator $\cO_j$ is as follows:  $\cO_j\Bigl(\sigma_j\Bigr)
= \lambda^{-\Delta_j} \cO_j\Bigl(\lambda \sigma_j\Bigr)$.} $\Delta_j$ equal
to 2, so that the integrals over ${\rm d}^2\sigma_j$ are conformaly invariant.
Appropriate operators include those present in \eq{ac}, such as:

\beq
\cO_G = G_{\mu\nu} \, :\diff_z \tilde{x}_{\mu} \,
\diff_{\bar{z}} \tilde{x}_{\nu}:\, . \label{operator}
\eeq
$\cO_G$ has a well defined conformal weight if $G_{\mu\nu} = f_{\mu\nu} \,
:\exp{\left[{\rm i} \, p_{\mu} \, \tilde{x}_{\mu}\right]}:$, where $f_{\mu\nu}$ is some polarization
from the target space point of view.

 In the integral \eq{massp} there is a region where $\sigma_1\to\sigma_2$
and close to it one could use the operator product expansion (OPE):

\beq
\lim_{\sigma_1\to\sigma_2} \, \cO_i(\sigma_1)\cO_j(\sigma_2) \approx \sum_k
C_{ijk} \, \left|\sigma_1 - \sigma_2\right|^{\Delta_k - \Delta_j - \Delta_i}
\,
\cO_k(\sigma_1),
\eeq
where the sum in the RHS runs over a basis of local operators in the world-sheet conformal
theory. Using this OPE, one finds \cite{Polyakov}:

\beqn
\cA_N = \int {\rm d}^2\eta\int \prod^N_{j = 2} {\rm d}^2\sigma_j \left\langle\cO_1(\sigma_2 + \eta)
\, \cO_2(\sigma_2) ... \cO_N(\sigma_N) \right\rangle \approx \nonumber \\ \approx \sum_k
C_{12k} \, \int^{a} {\rm d}^2\eta \, |\eta|^{\Delta_k - 4} \, \int \prod^N_{j=2}
{\rm d}^2\sigma_j \, \left\langle\cO_k(\sigma_2) \, \cO_3(\sigma_3) ... \cO_N(\sigma_N)\right\rangle +
\nonumber \\ + \, {\rm less \,\, singular \,\, terms} \approx \sum_k
\frac{1}{\Delta_k - 2} \cA^{(k)}_3 \, \cA^{(k)}_{N-1} + \, {\rm less \,\, singular
\,\, terms}, \label{ampl}
\eeqn
where $\cA^{(k)}_3 \propto C_{12k} \propto
\langle \cO_1 \, \cO_2 \, \cO_k \rangle$ and we have taken the integral over $|\eta|$ up to a scale
$a$ which is smaller than all other distances between $\sigma_j$'s.

  Now take into account that the operator $T(\tilde{x}) = \,
:\exp{\left[{\rm i} \, p_{\mu} \, \tilde{x}_{\mu}(\sigma)\right]}:$ has the conformal weight equal to
$\alpha' p^2_{\mu}/2$. One can find this weight using Wick's theorem for the
two-point correlation function  of this operator \cite{Polyakov}
and the propagator for $\tilde{x}_{\mu}$ from the action \eq{strac}.  Furthermore,
for $G$ proportional to $:\exp{\left[{\rm i} \, p_{\mu} \,
\tilde{x}_{\mu}(\sigma)\right]}:$ we have
the conformal weight for the operator \eq{operator} equal to
$\alpha'p^2_{\mu}/2 + 2$. Operators from \eq{ac} have the same conformal weight provided $B$
and ${\it \Phi}$ are proportional to $:\exp{\left[{\rm i} \, p_{\mu} \,
\tilde{x}_{\mu}(\sigma)\right]}:$. Likewise
for other sources (not present in \eq{ac}) also taken to be proportional to
$:\exp{\left[{\rm i} \, p_{\mu} \, \tilde{x}_{\mu}(\sigma)\right]}:$ one obtains $\alpha' p^2_{\mu}/2 +
\delta_k$, where $\delta_k > 2$ due to local operators which stand
near sources like the operator $:\diff_z\tilde{x}_{\mu} \,
\diff_{\bar{z}}\tilde{x}_{\mu}:$ which stands behind
$G_{\mu\nu}$ in eq. \eq{operator}.

  Thus, in any channel where $\sigma_i\to\sigma_j$ we have in the RHS of
\eq{ampl} a sum over all propagators of string excitations each
corresponding to some operator $\cO_k$:

\beq
\cA_N = \sum_k \frac{\cA^{(k)}_3 \, \cA^{(k)}_{N-1}}{p_{\mu}^2 + 2\, (\delta_k
- 2)/\alpha'}. \label{amplitude}
\eeq
In conclusion, there is a relation
between conformal weights of operators $\cO_k$ and masses of the
corresponding string states $m^2_k = 2\,(\delta_k - 2)/\alpha'$. Our
observation shows that $T$ describes a tachyonic state with $m_T^2 = -
p^2_{\mu} = - 4/\alpha'$, because $\delta_T = 0$.  At the same time
$G$, $B$ and ${\it \Phi}$ describe massless states
($\delta_{G,B,{\it \Phi}} = 2$), while
all other operators correspond to massive ones $(\delta_k > 2)$.

\subsection{A relation between gravity and string theory}

 Bearing the above considerations in mind, we could consider string theory at distances (set
by $G$, $B$, ${\it \Phi}$ and $T$) much bigger than $\sqrt{\alpha'}$.  First, in
this case one can replace separate quanta \eq{operator} by smooth
fields: as in the case of passing from photons to radio waves. Second,
in this situation massive string excitations are decoupled. This means that
we have to obtain a field rather than string theory at the distance scales in question.  In fact, a
free string is equivalent to infinitely many free particles: the string
propagator is just an infinite sum of particle propagators \eq{amplitude}.
Hence, forgetting about massive particles reduces the sum in \eq{amplitude} to
finitely many smallest mass excitations.

 In this way at scales in question and when $d=26$ one finds \cite{TsFr}:

\beqn
Z\left(\phantom{\frac12}G,B,{\it \Phi},T \, \, \right) = \frac{1}{16\pi \Gamma_N} \, \int
{\rm d}^{26}x \, \sqrt{-G} \, \exp{\left[-2{\it \, \Phi}\right]} \, \left[\cR \, +
\, 4 \, G^{\mu\nu} \, \diff_{\mu}{\it \Phi} \, \diff_{\nu}{\it \Phi} - \phantom{\frac12^{\frac12}} \right.
\nonumber \\ \left. \phantom{\frac12^{\frac12}} -
\frac{1}{12} \, H_{\mu\nu\lambda}^2 \, + \, \frac12 \, G^{\mu\nu} \, \diff_{\mu}T \, \diff_{\nu}T
\, + \, \frac12 \, m_T^2 \, T^2\right] \, + \, O(\alpha', \Gamma_N). \label{26}
\eeqn
This is 26-dimensional dilaton gravity interacting with the anti-symmetric
tensor $H_{\mu\nu\lambda} \propto \diff_{[\mu}B_{\nu\lambda]}$. Here $\Gamma_N
\propto g_{\rm s}^2 \alpha'^{12}$ is the 26-dimensional Newton's constant; from now on
$O(\alpha', \Gamma_N)$ {\it schematically} represents two-dimensional
$\sigma$-model and string loop corrections. If you will, the latter
contribution is due to string massive modes.

   Equation \eq{26} means that in the limit
$\alpha'\to 0$ (in comparison with a characteristic scale given by
functions $G,B,{\it \Phi}$ and $T$) the $Z$ functional gives exactly the same Feynman vertices and
propagators as the leading contribution in the RHS of \eq{26}.
Unfortunately, this fact can be explicitely established only for the
simplest background fields $G, B, {\it \Phi}$ and $T$. For example,
for the flat metric with constant background $B$ and ${\it \Phi}$.
Problems appear because there are no well developed methods of
quantization of the {\it non-linear} $\sigma$-model \eq{ac} with arbitrary sources $G, B, {\it \Phi}$ and $T$.
The best we can do now is to find that the vacua on the LHS and RHS of
\eq{26} are equivalent. In fact, conformal invariance of the $\sigma$-model \eq{ac}
imposes conditions on the sources \cite{Polyakov,GrScWi}: it is necessary to
have vanishing $\beta$-functions for the sources $G, B, {\it \Phi}$ and
$T$. These conditions are nothing but equations of motions for the action
\eq{26}.

  There is a way to intuitively understand why one should obtain
this particular action \eq{26} from string theory.  The action
\eq{ac} is invariant under infinitesimal transformations of $G$ and $B$
fields given by:

\beqn
G_{\mu\nu} \to G_{\mu\nu} + \diff_{(\mu} \xi_{\nu)}, \quad
B_{\mu\nu} \to B_{\mu\nu} + \diff_{[\mu} \rho_{\nu]}, \label{trans}
\eeqn
of which the first is nothing but the general covariance of the graviton
field. It is necessary (but not sufficient) to respect these invariances to
maintain the unitarity of the theory.  Now the goal would be to find a large
distance effective action for the sources in $Z$ which is invariant under the
transformations in question.  It is easy to see that action \eq{26} is
the only low energy one which obeys these conditions and includes
interactions with the dilaton ${\it \Phi}$.  The reason why ``$Z = S({\rm sources})$''
rather than ``$Z = \exp{\left[- \, {\rm i} S({\rm sources})\right]}$'' is that we are dealing with {\it first
quantized} string theory.

\subsection{Open bosonic string theory}

    Now let us consider open bosonic string theory. To maintain
Poincare invariance in the target space one naively (see the fourth
chapter) could think of using only the Neumann boundary conditions on the
coordinates $\tilde{x}_{\mu}$ of the open strings.

  As we have already mentioned, open string theory contains closed
string theory at loop level. Hence, open bosonic string theory contains all the
same sources in its generating functional as in \eq{str}, besides that, it
contains sources for its own excitations.  Furthermore
at open string ends one can add quantum (Chan-Paton) numbers (indices),
taking values in the fundamental representation of a gauge group.

  Thus, following the same reasoning as above one could find the massless
open string vertex operator to be a path-ordered Wilson exponent\footnote{Note
that there is also open string tachyon which we do not consider here.}:

\beq
{\rm tr \, P} \, \exp\left\{{\rm i} \int_{\rm boundary} \hat{a}_{\mu}(\tilde{x}) \, \diff_{\rm t}
\tilde{x}_{\mu} \, {\rm d}\tau \right\}, \label{wil}
\eeq
where $\diff_{\rm t}$ is a tangential derivative to the string's boundary
and $\tau$ is some parameterization of the latter. The presence of the
operator \eq{wil} in \eq{str} means that the ends of the strings are charged
with respect to $\hat{a}_{\mu}$. It is a gauge field, taking values in
the adjoint representation of the gauge group. In fact, the eq. \eq{wil} is
invariant under the gauge transformations: $\hat{a}_{\mu}\to \hat{a}_{\mu} +
\diff_{\mu}\hat{\lambda} + {\rm i} \left[\hat{a}_{\mu}, \hat{\lambda}\right]$.

   As for closed bosonic string theory, the open string generating
functional is equivalent at large distances to 26-dimensional dilaton gravity \eq{26}
interacting with YM theory for the gauge field $\hat{a}_{\mu}$.

\subsection{On unification of gravity and Yang-Mills theories}

 Let us discuss for a moment the possible relation of string theory to
quantized Einstein gravity and its unification with gauge interactions.  The
action \eq{26} is written through the use of the so called string metric.  From
the latter, one can pass to the standard Einstein metric through the
rescaling $G_E = G \, \exp{\left[- \, 4 \, {\it \Phi} \right]}$.  Hence, the
$26$-dimensional
Einstein-Hilbert action appears as a part of the large distance or classical
approximation to {\it quantum} string theory. Moreover, both gauge
and gravity theories could be treated on the same footing:  as approximations
to string theory. Furthermore one could obtain a four-dimensional theory via
so called compactifications \cite{GrScWi,Polchinski}. To do that one
considers $26$-dimensional world as a product of a non-compact
four-dimensional space with some very small compact $22$-dimensional one.
Both of the spaces in question should be solutions to the equations of motion following from \eq{26}.

     All that seems to be promising. However, closed bosonic string theory contains
a tachyonic excitation $T$ with $m_T^2 = - 4/\alpha'$.  This is a
pathological excitation. For its presence means that during the quantization
we have chosen an unstable vacuum. In fact, the tachyon is a negative mode
excitation over the vacuum, moreover, in closed bosonic string theory
higher self-interaction terms for the tachyon seemingly do not seal this
instability. Hence, one does not know the true vacuum or even whether it exists
at all in closed bosonic string theory.

\section{Type {\rm II} superstring theory}

   To obtain a sensible string theory one should consider supersymmetric
generalizations
of bosonic string theories \cite{GrScWi,WeBa}.
There are several non-anomalous types of superstring theories. Here we are going to discuss only closed
type {\rm II} strings in the Neveu-Schwarz-Ramond (NSR) formalism.  In this case
SUSY is added to bosonic string theory via anti-commuting $\psi_{\mu}$ fields which
are world-sheet super-partners of $\tilde{x}_{\mu}$. In principle one
must take into account the world-sheet metric field and its super-partner as well, however, as in
the case of bosonic string, via fixing symmetries of superstring theory we could
get rid of the fields in question.

   Thus, as a starting point we have $\cN=1$ two-dimensional
SUGRA interacting with conformaly invariant matter, represented by $\tilde{x}$ and
$\psi$ \cite{Polyakov,GrScWi,Polchinski}.  This is a SUSY extension of the
theory described by \eq{strac}.  Due to the presence of conformal symmetry the SUSY
reparametrization invariance of the action is enhanced to superconformal symmetry.  As we discuss
below, it is necessary to do some extra work to obtain SUSY inside target space.

 We consider here Hamiltonian quantization\footnote{Now that we have presented
the main statement of string theory, we would like to present
some other ways of quantizing the theory. For our purposes
Hamiltonian quantization \cite{GrScWi} of superstrings is more
transparent than the functional integral approach \cite{Polyakov}.} of
type {\rm II} superstring theories.  Free superstrings are described by the
action ($z = \exp{\left[\sigma_1 \, + \, {\rm i} \, \sigma_2\right]}$):

\beqn
S_{\rm sst} = \frac{1}{4\pi\alpha'} \, \int {\rm d}^2 z \, \left(\diff_z x^{\mu}
\, \diff_{\bar{z}} x_{\mu} + \bar{\psi}^{\mu} \, \diff_{\bar{z}}
\psi_{\mu} + {\rm c.c.} \right) + \nonumber \\ + \,
{\rm Faddeev-Popov \,\, ghost \,\, terms}, \label{supstr}
\eeqn
where we have eliminated the world-sheet metric field and its super-partner
via SUSY reparametrization and superconformal invariances.

  In theory \eq{supstr} one must impose the standard periodic
boundary conditions on $\tilde{x}_{\mu}$:  $\tilde{x}_{\mu}(\sigma_1,\sigma_2
+ 2\pi) = \tilde{x}_{\mu}(\sigma_1,\sigma_2)$.  At the same time, to respect
the aforementioned modular invariance, the quantum
theory of superstrings should contain sectors with two types of possible boundary
conditions for the world-sheet fermions \cite{Polyakov,GrScWi,Polchinski}.  The first
type of boundary condition is due to Ramond (R):

\beq
\psi_{\mu}(\sigma^2 + 2\pi) = \psi_{\mu}(\sigma^2),
\eeq
while the second one is due to Neveu and Schwarz (NS):

\beq
\psi_{\mu}(\sigma^2 + 2\pi) = - \psi_{\mu}(\sigma^2).
\eeq
with the same conditions for $\bar{\psi}_{\mu}$ in both cases. Therefore, there are two
kinds of
mode expansions for solutions of free two-dimensional Dirac
equation $\diff_{\bar{z}} \psi_{\mu} = 0$:

\beqn
\psi^{\mu}(z) = \psi_0^{\mu} + \sum_n \frac{b^{\mu}_n}{z^n}  \quad
({\rm R}), \nonumber\\ \psi^{\mu}(z) =
\sum_{n}\frac{c^{\mu}_{n+1/2}}{z^{n+1/2}} \quad ({\rm NS}).
\eeqn
There is a similar but independent expansion for $\bar{\psi}(\bar{z})$ as well.
It is conformal invariance which allows us to treat the left ($z$)
and right ($\bar{z}$) sectors independently: in conformal field theory they
do not interact with each other.

We omit the mode expansion for $\tilde{x}_{\mu}$ because the
corresponding creation operators do not lead to massless excitations
in superstring theory.

\subsection{Quantization and massless spectrum}

  To quantize superstring theory \eq{supstr} one imposes the
standard commutation (anti-commutation) relations on its bosonic (fermionic) fields.
Then the modes $d_n$ and $c_{n + 1/2}$ with positive and negative $n$'s
become, correspondingly, annihilation and creation operators.  At the same
time, the zero modes $\psi^{\mu}_0$ generate the algebra of Dirac
$\gamma$-matrices:

\beq
\left\{\psi^{\mu}_0, \psi^{\nu}_0\right\} = \eta^{\mu\nu}. \label{cl}
\eeq

  Superstring states are constructed by multiplications of
states from the left sector by states from the right sector that
satisfy a level matching condition.  Thus, since boundary conditions
can be assigned independently
in the left and right sectors, there are four kinds of states:

\beqn
\begin{array}{l l l}
{\rm NS} & - \quad \widetilde{\rm NS} \quad {\rm NS} & - \quad \tilde{\rm R} \, \\
{\rm R} & - \quad \widetilde{\rm NS} \quad {\rm R} & - \quad \tilde{\rm R} \, .
\end{array} \label{state}
\eeqn

 To find masses of excitations in this theory
it is necessary to use the two-dimensional energy-momentum tensor:

\beq
\cT(z) = \cT_{11} + \cT_{22} - 2 \, {\rm i} \, \cT_{12} = - \frac12 \, \left(\diff_z
\tilde{x}_{\mu}\right)^2 + \frac12 \, \psi_{\mu} \, \diff_z\psi_{\mu}
\label{tensorem}
\eeq
in the left sector. The energy-momentum tensor
in the right sector $\bar{\cT}(\bar{z})$ is the complex conjugate of \eq{tensorem}.
The corresponding conserved
Hamiltonians are $L_0 = \int {\rm d}z \cT(z)$ in the left sector and similarly
$\bar{L}_0$ in the right sector. Hence, the total Hamiltonian is $H = L_0 +
\bar{L}_0 + {\rm const}$, where the constant comes from the normal ordering and
has different values in the R- and NS-sectors \cite{GrScWi,Polchinski}.
With such a Hamiltonian one find the smallest mass sates \cite{GrScWi,Polchinski}:

\beqn
\begin{array}{l l l l}
{\rm mass | sector}  & & {\rm NS} & \qquad {\rm R} \\
m^2 = - 2/\alpha' & & \qquad \, |0 \rangle & \qquad - \\
m^2 = 0 & & c^{\mu}_{-1/2} \, |0\rangle  & \qquad |0\rangle, \,
\psi_0^{\mu} \, |0\rangle, \, \psi_0^{\mu}\, \psi_0^{\nu} \,
|0\rangle, \, ...
\end{array} \label{nsr}
\eeqn
and similarly in the $\widetilde{\rm NS}$- and
$\tilde{\rm R}$-sectors. The vacuum $|0 \rangle$ in
the R-sector is defined below, while $|0 \rangle$ in
the NS-sector is the standard vacuum for fermions.

  Furthermore to maintain the modular invariance one must project both the
left and right sectors to an eigen-state of the operator $(-1)^f$
\cite{Polyakov,GrScWi,Polchinski}. Here $f$ counts world-sheet fermion
number in superstring theory, i.e. this operator anti-commutes
with all fermionic creation and annihilation operators. That is to say that one must take the
partition function in superstring theory to be

$$
Z = {\rm  tr} \left\{\left[(-1)^{f^{\phantom{\frac12}}} \pm 1\right] \,
\exp{\left[- \, H\right]}\right\}
$$
with either plus or minus sign rather than just

$$
Z = {\rm tr} \left\{\exp{\left[- \, H^{\phantom{\frac12}}\right]}\right\}.
$$
This is the so called GSO projection. If one includes only those
states which obey

$$
\left[(-1)^f + 1\right]| {\rm state} \rangle = 0
$$
then the tachyon state $|0 \rangle$ in the NS-sector decouples from the spectrum, while
$c^{\mu}_{-1/2}|0 \rangle$ state survives\footnote{It is this kind of GSO
projection which leads, after taking account of the both left and right sectors,
to the appearance of SUSY in target space.  After the projection, the off-diagonal
elements in \eq{state} lead to the target-space superpartners for the
diagonal ones \cite{GrScWi,Polchinski}.}.  Thus, in the
NS--$\widetilde{\rm NS}$-sector we have $c^{\mu}_{-1/2}\bar{c}^{\nu}_{-1/2}|0,\bar{0} \rangle$ as the
massless state. Symmetric, anti-symmetric and trace part of which are related
to the familiar $G_{\mu\nu}$, $B_{\mu\nu}$ and ${\it \Phi}$ excitations in superstring theory.

  Let us now discuss what happens in the left R-sector (consideration of the
right $\tilde{R}$ sector is similar) \cite{Polchinski}.  We change
the basis of the zero modes $\psi^{\mu}_0$ to

\beqn
d^{\pm}_0 = \frac{1}{\sqrt{2}}\left(\psi^1_0 \mp {\rm i} \, \psi^0_0\right)
\qquad d^{\pm}_i = \frac{1}{\sqrt{2 {\rm i}}}\left(\psi^{2i}_0 \pm
\psi^{2i+1}_0 \right), \quad i = 1,...,4.
\eeqn
Then from \eq{cl} one gets:

\beq
\left\{d^+_I, d^-_J\right\} = \delta_{IJ}, \quad I,J = 0,...,4.
\eeq
These $d^{\pm}_I$ generate $2^5 = 32$ Ramond ground states
$|s \rangle = |\pm 1/2 , ... , \pm 1/2 \rangle$ as:

\beqn
d^{\pm}_I\left|\pm \frac12,..., \pm \frac12 \right\rangle = 0, \nonumber \\ d^+_I
\left|-\frac12,...,s_I=-\frac12,...,-\frac12 \right\rangle =
\left|-\frac12,...,s_I=+\frac12,...,-\frac12 \right\rangle \label{loenst}
\eeqn

   One can verify that after a SUSY reparametrization gauge fixing there
are Super-Virasoro conditions on the physical states of superstring theory
\cite{GrScWi,Polchinski}. They appear as the standard conditions of Dirac's
approach to Hamiltonian quantization. These conditions are:

\beqn
\cT \, |{\rm state} \rangle = 0, \quad \bar{\cT} \, |{\rm state} \rangle = 0 \nonumber \\
\diff_{z} \tilde{x}_{\mu} \, \psi_{\mu} \, |{\rm state} \rangle = 0, \quad
\diff_{\bar{z}}\tilde{x}_{\mu} \, \bar{\psi}_{\mu} \, |{\rm state}\rangle = 0,
\label{Virasoro}
\eeqn
which are nothing but the conditions of superconformal invariance of superstring theory. To cancel
the anomaly in this case one has to take $d=10$ rather than $d=26$.

  Now from the first condition in the second row of \eq{Virasoro} it follows
that:  $p_{\mu}\, \psi^{\mu}_0 \, |{\rm state} \rangle = 0$. At the same time, in the reference
frame where $p^{\mu} = (p^0,$ $p^0,0,...,0)$ one has that
$p_{\mu}\psi^{\mu}_0 = \sqrt{2}p^0 d^+_0$.  Hence, $s_0 = + 1/2$, which
leaves only $s_i = \pm 1/2, \,\, i = 1,...,4$, i.e. $16$ physical vacua:
$8_s$ with an even number of $\left(-1/2\right)$ and $8_c$ with an odd
number of $\left(-1/2\right)$. These $8_s$ and $8_c$ states compose different
chirality spinor representations of the ten-dimensional Lorenz
group \cite{GrScWi,Polchinski}. In fact, $\psi^0_{\mu}$
generate the algebra of ten-dimensional Dirac matrices \eq{cl} and $8_c$ and $8_s$
are its two irreducible representations.

The GSO projection keeps one of these states ($8_c$ or $8_s$) and removes
the other.  Taking into account that there are two possibilities for the
vacuum:

\beq
(-1)^f\left|-\frac12,...,-\frac12 \right\rangle = \pm \left|-\frac12,...,-\frac12\right\rangle,
\label{choice}
\eeq
one concludes that there can be two types of theories.
If we choose opposite signs for the vacua in the
R- and $\tilde{\rm R}$-sectors, we obtain non-chiral type {\rm II}A
theory.  If we choose the same sign, then we have chiral type {\rm II}B theory.

  In conclusion, in the R and $\tilde{\rm R}$-sectors the massless states \eq{loenst} have
target space fermionic quantum numbers \cite{GrScWi,Polchinski}: depending
on a choice \eq{choice} they are ten-dimensional fermions ether of one chirality $|\beta\rangle$ or
of the other $|\dot{\beta}\rangle$.  Schematically it means that in the
R--$\tilde{\rm R}$ sector there are states as follows:

\beqn
{\rm (IIA)} \quad
\left(\gamma^{[\mu_1}...
\gamma^{\mu_{n+1}]}\right)_{\lambda \dot{\beta}}|\lambda\dot{\beta}\rangle,
\nonumber \\ {\rm (IIB)} \quad
\left(\gamma^{[\mu_1}...
\gamma^{\mu_{n+1}]}\right)_{\lambda\beta}|\lambda\beta\rangle, \label{r-r}
\eeqn
where $\gamma_{\mu}$ are ten-dimensional Dirac matrices in the Weyl-Majorana
representation.  These states correspond to the bosonic tensor fields
$A_{\mu_1...\mu_n}$ with the field strengths $F_{\mu_1...\mu_{n+1}} =
\diff_{[\mu_{n+1}} A_{\mu_1...\mu_n]}$. Now we see that due to the chirality
properties of the massless states, in type {\rm II}A theory there are
{\it only odd} rank $A$ fields. At the same time in type {\rm II}B theory {\it only even} rank
$A$ fields are present.

  Type {\rm II} string theories are invariant under {\it two} SUSY transformations in the target
space ($Q$ and $\tilde{Q}$), which correspond to the left and right sectors
on the world-sheet, respectively \cite{GrScWi,Polchinski}. This is the reason
why one refers to these string theories as type {\rm II}.

\subsection{Type {\rm II}B superstrings at large distances}

   Below we mostly consider type {\rm II}B
string theory (type {\rm II}A theory is very similar) with its bosonic
massless excitations. Besides the standard NS--$\widetilde{\rm NS}$ fields $G,B$ and
${\it \Phi}$, this theory
contains R--$\tilde{\rm R}$ fields which are the scalar $A$, two-form tensor
potential $A_{\mu\nu}$, four-form tensor potential $A_{\mu\nu\alpha\beta}$
and their duals.  In fact, by construction, among the fields described in
\eq{nsr}-\eq{r-r} there are various duality relations:

\beqn
F_{\mu_1...\mu_9} &=& \epsilon_{\mu_1...\mu_{10}} \,
\diff_{\mu_{10}}A \nonumber\\
F_{\mu_1...\mu_7} &=& \epsilon_{\mu_1...\mu_{10}} \,
F^{\mu_8\mu_9\mu_{10}} \nonumber\\
F_{\mu_1...\mu_5} &=& \epsilon_{\mu_1...\mu_{10}} \,
F^{\mu_6...\mu_{10}}. \label{sd'}
\eeqn
Here $\epsilon_{\mu_1...\mu_{10}}$ is the completely anti-symmetric
tensor in ten dimensions.

 As in bosonic string theory, superstrings contain target space
SUGRA at large distances. In the case of superstring theory it is known how to
calculate its generating functional only if $d = 10$, which is when the
superconformal anomaly is canceled.  Thus, the bosonic part of the large
distance type {\rm II}B ten-dimensional SUGRA action is
\cite{GrScWi,Polchinski}:

\beqn
S_{\rm IIB} = \frac{1}{16\pi\Gamma_N} \, \int {\rm d}^{10}
x \, \sqrt{-G} \, \left\{\exp{\left[- \, 2 \, {\it \Phi}\right]} \, \left[\phantom{\frac12^{\frac12}} \cR
 \, \, + \, \,
4 \, G^{\mu\nu} \, \diff_{\mu}{\it \Phi} \, \diff_{\nu}{\it \Phi}
- \frac{1}{12} \, H^2_{\mu\nu\gamma} \, \, \,
\right] \, + \right. \nonumber\\ - \frac12 \, \left[G^{\mu\nu}\diff_{\mu} \, A \diff_{\nu} \, A +
\tilde{F}^2_{\mu\nu\gamma}
+ \frac12 \, \tilde{F}^2_{\mu_1...\mu_5} \, + \phantom{\frac12^{\frac12}}
\right. \nonumber \\ \left.\left.
+ \phantom{\frac12^{\frac12}} \frac12 \,
\epsilon_{\mu_1...\mu_{10}} \, A_{\mu_1...\mu_4} \,
H_{\mu_5\mu_6\mu_7} \, F_{\mu_8\mu_9\mu_{10}}\right]\right\} + {\rm fermions} +
O(\alpha',\Gamma_N) \nonumber \\ {\rm where} \quad \tilde{F}_{\mu\nu\gamma} =
F_{\mu\nu\gamma} - A \, H_{\mu\nu\gamma} \quad {\rm and}  \nonumber \\
\tilde{F}_{\mu_1...\mu_5} = F_{\mu_1...\mu_5} - \frac12 \, A_{[\mu_1\mu_2}
\, H_{\mu_3\mu_4\mu_5]} + \frac12 \, B_{[\mu_1\mu_2} \, F_{\mu_3\mu_4\mu_5]}.
\label{2b}
\eeqn
Here $\Gamma_N = 8\pi^6 g_{\rm s}^2\alpha'^4$ is the ten-dimensional Newton's
constant.  Furthermore in this action one must impose the self-duality
condition on the R--$\tilde{R}$ four-form field as shown in the last row of
\eq{sd'}.  There are also various dual versions of type {\rm II}B
SUGRA, which are expressed through the dual tensor fields from \eq{sd'}.

   Thus, we see that superstring theories are self-consistent and lead
at large distances (classical limit) to SUGRA theories. Now we are ready to discuss various
solitons in SUGRA and string theories.

\section{D-branes and SUGRA solitons}

  In SUGRA theory there are many different
solitons \cite{Polchinski,Duff}. In ten dimensions they could be
particle-like black holes or different types of branes (membranes etc.) which
are multi-dimensional analogues of four-dimensional black holes. Their singularities live on
multi-dimensional sub-manifolds of ten-dimensional target space and are
surrounded by multi-dimensional event horizons. They can be neutral or
charged with respect to some tensor gauge fields (like $B_{\mu\nu}$ or the
R--$\tilde{R}$ fields discussed in the previous section) just as
point-like black-holes can be charged with respect to gauge {\it
vector} fields: one could surround the locus of a soliton
by a multi-dimensional sphere and then find a flux of a corresponding
tensor field.

  Now, keeping in mind the fact that string theory suggests us
quantization of gravity, one could ask what are the quantum counterparts
of these solitons? Besides being of academic interest, the answer to this
question could reveal some features of black hole
thermodynamics \cite{Polchinski,Malbl}. Furthermore, as we discuss
below, it gives a relation between SYM theory and SUGRA.

   The problem is that to pass from large distance gravity to microscopic
string theory one needs to vary the parametrs $\alpha'$ (measured
with respect to a characteristic scale) and $g_{\rm s}$ in string theory. It happens
that during this variation when background fields are turned on, corrections
$O(\alpha',\Gamma_N)$ in \eq{2b} could become more relevant than the leading
large distance contribution. As we briefly discussed in the introduction, these
corrections even could change the form of the background completely. First, this
destroys the event horizon, which appears to be a low
energy global characteristic \cite{Polchinski,Maldacena}.  Geometrically it is
seen when the size of the horizon of a soliton becomes smaller than the string characteristic scale.
Second, a variation of the parameters in question could lead to an uncontrollable renormalization of the charge and
tension of a soliton or even to a change of the fundamental degrees of
freedom in the theory. In fact, we do not have the complete knowledge of string theory
dynamics.

   However, in the presence of SUSY one could control the renormalization of the
low energy (large distance) action.  Furthermore there are solitons
in SUSY theories for which the renormalizations of
their mass and charge are under control \cite{WiOl}. They are referred to as
BPS solitons and respect at least some part of the SUSY transformations in
such theories. Note that arbitrary
excitations do not respect any symmetries, while the fact that
SUSY is respected imposes strong restrictions on possible dynamics \cite{WeBa}.

  Furthermore of all BPS solitons in string theory, quantum
counterparts are known only for those which are charged with respect to the
R--$\tilde{\rm R}$ tensor fields. For only in the latter case does a good
two-dimensional conformal field theory description exist.  Although
historically R--$\tilde{\rm R}$ BPS SUGRA solitons were found first \cite{Duff} and only
after that their quantum D-brane
description \cite{PolD}, we start our discussion with the definition of
D-branes.  Then we explain their relation to SUGRA solitons and to
SYM theory.

\subsection{Definition of D-branes}

  One could wonder if it is possible to consider open string sectors in
closed type {\rm II} superstring theories.  It appears that to avoid anomalies
\cite{Polchinski} open strings in these sectors should have both Neumann (N) and
Dirichlet (D) type boundary conditions on string coordinates
\cite{PolD}:

\beqn
\diff_{\rm n} x_m = 0 , \quad \psi^m = \pm \tilde{\psi}^m,
\quad m = 0,...,p \quad ({\rm N})\nonumber\\
x^i = C^i, \quad \psi^i = \mp \tilde{\psi}^i \quad i = p+1,...,9 \quad
({\rm D}), \label{bou}
\eeqn
where $C^i$ are some fixed numbers and $\diff_{\rm n}$ is a normal
derivative to the string boundary.  Therefore, in such a situation the ends of
open strings can freely move only along directions labeled by ``$m$''. In fact, they are
confined to $(p+1)$-dimensional sub-manifolds placed at $x_i = C_i$ in
ten-dimensional target space.  These sub-manifolds, filling completely ``$p$''
directions and situated at $x_i = C_i$, are referred to as Dp-branes.  At the
same time in the bulk of the target space there are ordinary type {\rm II}
closed strings.

   The Dp-branes have several features which are relevant for our further
discussion.  First, they break Poincare invariance inside target space
${\rm P(10) \to P(1+p)\times SO(9-p)}$.  Hence, to maintain P(10), one should consider these
Dp-branes as dynamical excitations in superstring theory.
Second, to respect SUSY one must consider $p=0,2,4,6,8$ for type
{\rm II}A and $p=-1,1,3,5,7$ for type {\rm II}B theories\footnote{The case $p=-1$ describes
the so called D-instanton, which is a D-brane whose ``world-volume'' is just a
point in the ten-dimensional Euclidean target space. This D-instanton is described by
open strings with Dirichlet type boundary conditions in all
ten directions.} \cite{Polchinski} (see
below). Third, because of the boundary conditions \eq{bou}, the Dp-branes
can not respect more than half of SUSY transformations in type {\rm II} string
theories.  In fact, the two SUSY transformations (due to $Q$ and $\tilde{Q}$)
are related to each other:  the left and right sectors on the string
world-sheets are no longer independent because of the boundary conditions.

     Interactions of a Dp-brane with the massless closed
string excitations are described by \cite{PolD}:

\beqn
Z\left(\phantom{\frac12}G,B,{\it \Phi}, \{A\}, a, \phi, {\rm fermions} \,\,\right) = \sum^{\infty}_{\cG = 0} \, \int
[\cD h_{ab}]_{\cG} \, \cD \tilde{x}_{\mu} \, \cD \left({\rm fermions}\right) \times \nonumber \\
\times \exp{\left[- \, {\rm i} S_{\rm Dst}\left(\tilde{x}_{\mu}^{\phantom{\frac12}}, h_{ab}, G_{\mu\nu}, B_{\mu\nu},
{\it \Phi}, \{A\}, a, \phi, {\rm fermions}\right) \right]} \nonumber \\ {\rm where} \quad
S_{\rm Dst}\left(\tilde{x}_{\mu}^{\phantom{\frac12}}, h_{ab}, G_{\mu\nu}, B_{\mu\nu},
{\it \Phi}, \{A\}, a,
\phi, {\rm fermions}\right) = \nonumber \\ =
\frac{1}{2\pi\alpha'} \, \int
{\rm d}^2\sigma \, \left\{\sqrt{-h} \, h^{ab} \, G_{\mu\nu}(\tilde{x}) \, \diff_a
\tilde{x}^{\mu} \, \diff_b \tilde{x}^{\nu} + \phantom{\frac12^{\frac12}} \right. \nonumber\\ \left.
\phantom{\frac12^{\frac12}} + \, \epsilon^{ab} \,
B_{\mu\nu}(\tilde{x}) \, \diff_a \tilde{x}^{\mu} \, \diff_b \tilde{x}^{\nu} +
\alpha' \, \sqrt{-h} \, R^{(2)} \, {\it \Phi}(\tilde{x}) \right\} + {\rm R-\tilde{R}} \,\, {\rm fields}
+ \nonumber \\ + \int
{\rm d}\tau \, a_m(\tilde{x}_m) \, \diff_{{\rm t}} \tilde{x}_m + \int
{\rm d}\tau \, \phi_i(\tilde{x}_m) \, \diff_{{\rm n}} \tilde{x}_i +
{\rm fermions}. \label{sup}
\eeqn
Here $\tau$ is some parametrization of the boundary.
As usual, closed strings appear at the loop level inside open string theory.
In this functional we have fixed a light-cone gauge: $\phi_m (\tilde{x}) = \tilde{x}_m$,
where $\phi_{\mu} = (\phi_m, \phi_i)$ describes the embedding of the Dp-brane
into target space.

  Let us clarify the meaning of the quantity \eq{sup}.
If one puts $G_{\mu\nu} = \eta_{\mu\nu}$, $B_{\mu\nu} = 0$, ${\it \Phi} = 0$
and all R--$\tilde{\rm R}$ fields with fermions to zero then equation \eq{sup} describes the time evolution of a quantum state
in a two-dimensional conformal field theory. In fact, taking a time slice we fix initial conditions
and the boundary conditions
\eq{bou} and integrate over all fields in the theory with these boundary and initial conditions.
This is by definition a quantum state. Adding time gives us the time evolution of this state.
In the case when all background fields are non-trivial, the quantity \eq{sup}
describes interactions of the quantum state with these fields. Moreover,
we show below that \eq{sup} at large distances describes the interactions
of a SUGRA soliton --- classical
limit of the quantum state in question --- with the
aforementioned SUGRA fields.

  Let us explain, following \cite{Wit95}, the origin of the sources $a_{m}$
and $\phi_i$ in \eq{sup}.  As we have already noted, string
theory should be invariant under the transformations described by \eq{trans}. For a
closed string they are respected, but when a string world-sheet
has a boundary there are boundary terms appearing after such transformations. To
cancel the first of the transformations in \eq{trans}, one must add a field $\phi_i$ at
the string boundary.  It should transform as $\phi_i \to \phi_i -
\xi_i/\alpha'$  to compensate \eq{trans}.  While the boundary
term appearing as a result of the transformation \eq{trans} along the Dp-brane
vanishes, because Poincare invariance is respected there.  Hence, $\phi_i$
appear as pure gauge degrees of freedom would if there were no
breaking of Poincare invariance in the presence of a Dp-brane.
Furthermore from this consideration it is clear what the physical meaning
of these fields is: They represent transverse fluctuations of the Dp-branes
around their positions $C_i$.  That is to say $C_i$ are just VEV's of the fields
$\phi_i$: $\phi_i + C_i \to \phi_i$.

  Likewise, to maintain the second invariance in \eq{trans}, the string boundaries
should be charged with respect to an Abelian gauge field $a_{m}$. In this
case the boundary term appearing after the second transformation \eq{trans} is compensated by a
shift $a_{m}\to a_{m} - \rho_m/\alpha'$. (This shift is
different from the ordinary gauge transformation $a_{m} + \diff_{m}\lambda$
of the field $a_{m}$.) The physical meaning of the fields $a_m$ is that
they describe longitudinal fluctuations of the Dp-branes.

\subsection{D-branes at low energies}

  At energies much smaller than $1/\sqrt{\alpha'}$ the functional
\eq{sup} acquires the following form \cite{Lei89}:

\beqn
Z\left(\phantom{\frac12} G,B,{\it \Phi}, \{A\}, a, \phi, {\rm fermions}  \, \, \right) =
S_{\rm II}\left(\phantom{\frac12} G,B,{\it \Phi}, \{A\}, {\rm fermions} \,\, \right) + \nonumber \\ +
m_p \, \int {\rm d}^{p+1} x \, \exp{\left[- \, {\it \Phi}\right]} \, \sqrt{-\det\left(g_{mn} + b_{mn} +
2 \, \pi \, \alpha' \, f_{mn}\right)} + \nonumber \\ + Q_p \, \int {\rm d}^{p+1}x
\, \epsilon_{0...p} \, A_{0...p} + {\rm fermions} + O(\alpha',\Gamma_N),\label{BI}
\eeqn
where $\epsilon_{0...p}$ is the $0...p$ component of the
$(p+1)$-dimensional totally anti-symmetric tensor, and

\beqn
f_{mn} = \diff_{[m} a_{n]},
\quad g_{mn} &=& G_{ij} \, \diff_m \phi_i \, \diff_n \phi_j + G_{i(m}\diff_{n)}\phi_j +
G_{mn}, \nonumber \\ b_{mn} &=& B_{ij} \, \diff_m\phi_i \, \diff_n\phi_j +
B_{i[m}\diff_{n]}\phi_j + B_{mn}.
\eeqn
are field strength for $a_m$, the induced metric, and the $B$ field on the Dp-brane
world-volume. Note that in the action
\eq{BI} we maintain all powers of $f_{mn}$ while neglecting its derivatives.

  In eq. \eq{BI} $S_{\rm II}$ is the type {\rm II} ten-dimensional
($S_{\rm II} \propto \int {\rm d}^{10} x ...$) SUGRA action. In the case of type
{\rm II}B
string theory $S_{\rm II}$ is given by the leading contribution in \eq{2b}. The
second contribution in \eq{BI} is so called Dirac-Born-Infield (DBI)
action for non-linear $(p+1)$-dimensional ($\propto \int {\rm d}^{p+1} x ...$)
electrodynamics.  Its coefficient is the mass per unit volume of the Dp-brane
and can be found to be equal \cite{Polchinski} to

$$
m_p = \frac{\pi}{g_{\rm s}}
\left(4\pi^2\alpha'\right)^{-(p+1)/2}.
$$
The third term shows that Dp-branes are sources for
the R--$\tilde{\rm R}$ tensor fields $A$.  In another words
Dp-branes are charged with respect to the $(p+1)$-tensor
R--$\tilde{\rm R}$ fields
with charges equal to $Q_p$ \cite{PolD}.  Taking into account the special
properties of R--$\tilde{\rm R}$ fields (discussed following equation \eq{r-r}), it
is clear why there could be only Dp-branes with $p=0,2,4,6,8$ in type
{\rm II}A and $p=-1,1,3,5,7$ in type {\rm II}B theories \cite{Polchinski}.

    What is important for our further discussion is that the action \eq{BI}
is SUSY invariant. In fact, Dp-branes \eq{sup} respect half of the SUSY invariance in type
{\rm II} string theories, and obey

\beq
m_p = Q_p l^{-(p+1)}_{\rm st} \quad {\rm where} \quad l_{\rm st} \sim \sqrt{\alpha'}. \label{cancel}
\eeq
The force between any two equivalent and parallel Dp-branes vanishes
\cite{Polchinski}. This is because the
repulsion due to the R--$\tilde{\rm R}$ tensor field compensates the gravitational
attraction. This is called the ``No force condition'' and is
important for our further considerations.

\subsection{D-branes as sources for {\rm R}--$\tilde{\rm R}$ SUGRA
solitons}

   Now let us probe a Dp-brane at large distances, when $r = \sqrt{x_i
x^i} \gg \sqrt{\alpha'}$ (note that $g_{\rm s}\to 0$ to suppress the
corrections $O(\alpha',\Gamma_N)$).  We reiterate that it is the SUSY
invariance of the action \eq{BI} which allows us to go easily from large to
small $r$ (and vice versa) and the leading contribution in \eq{BI} does not
change. Hence, in the process of going to large $r$ we could just forget
about the Dp-brane excitations $a_m$ and $\phi_i$.  We mean that a large
distance observer does not feel them and one could substitute in \eq{BI}
their classical values: $a_m = \phi_i = 0$ if there are no sources for these
fields. Thus, if there are no non-trivial background fields $G$, $B$ and
${\it \Phi}$, we have:

\beq
Z = S_{\rm II} + m_p\int {\rm d}^{p+1} x + Q_p \int {\rm d}^{p+1} x \,
\epsilon_{0...p} \, A_{0...p} +  O(\alpha',\Gamma_N), \label{Dsur}
\eeq
The second and the third terms in this equation are just sources for the
curvature and the corresponding R--$\tilde{\rm R}$ field.  They could be
rewritten as $\int {\rm d}^{p+1} x ... \propto \int {\rm d}^{10} x
\delta^{9-p}(x_i - C_i) ...\,$, so solutions of the classical
equations of motion for \eq{Dsur} with these sources appear to be
BPS R--$\tilde{\rm R}$ SUGRA solitons:

\beqn
{\rm d}s^2 = f^{-1/2}_p \, {\rm d}x_m {\rm d}x^m  +
f^{1/2}_p \, \left({\rm d}r^2 + r^2 {\rm d}\Omega^2_{8-p}\right), \nonumber\\
\exp{\left[- \, 2 \, {\it \Phi}\right]} = f_p^{(p-3)/2}, \nonumber\\
A_{0...p} = - \frac12 \, \left(f^{-1}_p - 1 \right), \label{sol}
\eeqn
where $p=0,2,4,6,8$ in type {\rm II}A and $p=-1,1,3,5,7$ in type {\rm II}B
theories \cite{Duff}. These solutions are states in SUGRA
which are classical limits of the states \eq{sup} in string theory.
This how one finds a relation between the Dp-branes and
R--$\tilde{\rm R}$ Dp-brane SUGRA solitons. At the same time, \eq{Dsur}
describes low energy fluctuations around these SUGRA solutions.

  All the solutions \eq{sol} are BPS and for any function $f_p$ they
preserve half of the SUSY transformations in SUGRA theory.
The equations of motion of SUGRA theory \eq{Dsur} (related to the closure of the SUSY algebra)
imply \cite{Duff} that $f_p$ should obey

\beq
\Delta^{9-p} f_p(r) = m_p \cdot \delta^{9-p}\left(x_i - C_i\right).
\label{harms}
\eeq
Here $\Delta^{9-p}$ is the Laplacian for the flat metric in the
directions $p+1,...,9$.  Hence, one gets:

\beq
f_p = 1 + \left(\frac{r_p}{r}\right)^{7-p}, \label{f}
\eeq
where $r_p \propto 1/m_p^{1/(p+1)}$. Note that one can neglect string
corrections to \eq{Dsur} in the case $r_p \gg \sqrt{\alpha'}$.

  We now consider $N$ Dp-branes parallel to each other and placed at
$\vec{r}_s$, $s = 1,..., N$. We can do that safely because of the ``No force condition''.
At low energies (large distances) such a system of $N$ Dp-branes corresponds
to a R--$\tilde{\rm R}$ Dp-brane soliton \eq{sol} with the charge $Q_p \propto N$
and

\beq
f_p = 1 + \sum^N_{s=1}\left(\frac{r_p}{|\vec{r} - \vec{r_s}|}\right)^{7-p}.
\eeq
The tension of the soliton is $M_p = N m_p$. Note that when one puts all
the Dp-branes on top of each other ($r_s = 0$ for all $s$) then

\beq
f_p = 1 + \left(\frac{R_p}{r}\right)^{7-p}, \label{fg}
\eeq
where $R_p^{7-p} = N r_p^{7-p}$. In this case one could neglect string theory
corrections to \eq{Dsur},\eq{sol},\eq{fg} if $R_p \gg \sqrt{\alpha'}$.

 The solitons \eq{sol} are multi-dimensional analogs of the
four-dimensional critical Reissner-Nordstrom black hole.
Note that the event horizon of these solutions is at $r=0$.

\subsection{D-branes and SYM}

  Now let us probe a Dp-brane at small distances $r \ll R_p$ with $g_{\rm s} \to 0$. In this case one could forget
about long wave-length fluctuations of the bulk fields $G$, $B$, ${\it \Phi}$ and
$\{A\}$. Hence, these fields are equal to their classical values, i.e.  to zero in
the absence of external sources. Thus, expanding \eq{BI} in powers of a small
$f_{mn}$, one obtains:

\beqn
Z = S_{\rm II} + S_{\rm SQED} + O(\alpha',\Gamma_N), \quad {\rm where} \nonumber \\
S_{\rm SQED} \propto \int {\rm d}^{p+1} x \, \left[ \frac12 f^2_{mn} + \frac12
|\diff_m\phi^i|^2 + ...\right]. \label{bibi}
\eeqn
Dots in the second row stand for the fermionic super-partner terms. The latter could be
recovered from the fact that this supersymmetric QED (SQED) is maximally supersymmetric
in $(p+1)$ dimensions. In fact, we know from \eq{bou},\eq{sup} the number of SUSY
transformations under which the theory \eq{bibi} is invariant. This number is 16
--- half of 32, which is the total number of components of supercharges
in type {\rm II} string theories.

  There is also another way to find the number of SUSY transformations under which
\eq{bibi} is invariant. One could consider ten-dimensional $\cN=1$ (maximally
supersymmetric: there are
16 components of supercharges) SQED:

\beq
L = \frac12 f^2_{\mu\nu} + \frac{\rm i}{2}\bar{\Psi} \hat{\diff} \Psi
\eeq
where $\Psi$ are Majorana-Weyl spinors and the super-partners of $a_{\mu}$.
Then one can make a reduction of the theory to $(p+1)$ dimensions.
That is when one considers all the fields in the theory
to be independent of $(9-p)$ coordinates \cite{WeBa}.
This way, changing notation from $a_i$ to $\phi_i$ ($i=p+1,...,9$), one gets
the theory \eq{bibi} with the proper fermionic content. Furthermore during
this procedure the number of SUSYs is increased with respect to $\cN = 1$ in
ten dimensions \cite{WeBa}.  In fact, the ten-dimensional fermions $\Psi$ are
rearranged into representations of the smaller Poincare group P(p+1).
Hence, from a single ten-dimensional fermion we obtain several lower-dimensional ones.

   The low energy action \eq{bibi} could also be found from
another point of view \cite{Wit95}. At low energies strings which
terminate on Dp-branes look like massless vector ($a_m$) and scalar
($\phi_i$) excitations --- the massless excitations in open string theory
\cite{GrScWi}. Furthermore in the limit $g_{\rm s}\to 0$ the coupling of open
strings attached to Dp-branes with closed strings in the bulk is suppressed.
At this point one finds that the low energy theory for such excitations is
SUSY QED --- the only supersymmetric and gauge-invariant action containing
smallest number of powers of derivatives of the fields.

   The last point of view is helpful in understanding the low energy theory
describing a bound state of Dp-branes \cite{Wit95}. Let us consider $N$
parallel Dp-branes with the same $p$. In this situation in addition to the strings
which terminate on the same Dp-brane, there are strings stretched
between different branes. Furthermore because the strings are
oriented, there could be two types of
strings stretched between any two Dp-branes. The strings attached with both ends to the same
Dp-brane give
familiar massless vector excitations living on the brane.  On the other hand,
the stretched strings give vectors with masses proportional to distances
between corresponding Dp-branes. They are charged with respect to the gauge
fields living on the Dp-branes at their ends. Therefore, the latter vector
excitations are similar to the $W^{\pm}$-bosons in gauge theories with
spontaneous symmetry breaking.  They acquire masses through a kind of Higgs
mechanism --- splitting of Dp-branes --- and become massless when the Dp-branes
approach each other.  Hence, the world-volume theory on the
bound state of $N$ Dp-branes is nothing but U$(N)$ maximally supersymmetric
SYM theory \cite{Wit95}:

\beqn
S \propto M_p \, \alpha'^2 \, \int {\rm d}^{p+1}x \, {\rm tr} \left\{\hat{f}_{mn}^2 +
\left|D_m\hat{\phi}_i\right|^2 + \sum_{i>j}\left[\hat{\phi}_i,
\hat{\phi}_j\right]^2 + ... \right\}, \nonumber \\ {\rm where} \quad
\hat{f}_{mn} = \diff_{[m} \hat{a}_{n]} + {\rm i} \left[\hat{a}_m, \hat{a}_n\right]
\quad {\rm and} \quad D_m = \diff_m + {\rm i}\left[\hat{a}_m, \cdot\right].
\label{SYM}
\eeqn
Dots in this action stand for fermionic terms. Furthermore all
possible positions of the Dp-branes, composing this bound state, are given by
VEV's of the ${\rm U(N)}$ matrix $\hat{\phi}_i$. Note that the potential
in the action \eq{SYM} has flat directions. These flat directions are not lifted
by quantum corrections due to the SUSY invariance of the action \eq{SYM}.
Thus, the ${\rm U(1)}$ factor in the decomposition
${\rm U(N) = SU(N) \times U(1)}$ describes the center
of mass position of the Dp-brane bound state. Unfortunately we do not know any rigorous derivation of \eq{SYM} from
first principles such as the definition of the Dp-branes, though, there is a
non-canonical way to formulate the non-Abelian version of \eq{sup}, \eq{BI},
and hence of \eq{bibi} \cite{Tseytlin}, which could be useful for the
derivation of \eq{SYM}.

  Anyway, to sharpen the reader's understanding we give one more argument
in favor of the appearance of SYM on the Dp-branes.  When one has a stack
of Dp-branes, the strings which terminate on them carry Chan-Paton indexes,
enumerating these Dp-branes. Hence, one would obtain sources like \eq{wil} for
their massless excitations, where $\hat{a}_i \to \hat{\phi}_i, \,\, i=p+1,...,9$. This, as we
know, leads at large distances to SYM theory and shows that the theory \eq{SYM}
is the reduction of ten-dimensional $\cN = 1$ SYM to $(p+1)$ dimensions.

  It is worth mentioning at this point that one could also consider BPS
bound states of different types of Dp-branes (with different p's)
\cite{Polchinski}.  However that is outside the scope of this discussion.

\section{AdS/CFT-correspondence}

   We see that the Dp-branes have two different
descriptions depending on what distance one looks at them. From far
away the D-branes look like sources for gravity solitons, while at small
distances one sees their quantum fluctuations described by SYM
theory.  It seems that both limits are unrelated to each other, however,
this is not so.  To understand why, from now on we are going to
discuss one of the simplest situations.

  We consider a stack of $N$ D3-branes in ten-dimensional type {\rm II}B SUGRA.
The D3-branes are on top of each other at $x_4 = ... = x_9 = 0$ and
occupy $0,...,3$ directions. The corresponding SUGRA soliton is the self-dual
R--$\tilde{\rm R}$ D3-brane \eq{sd'}, \eq{sol}, \eq{fg} with

\beq
R_3^4 = 4\pi g_{\rm s} N \alpha'^2 \quad {\rm and} \quad Q_3 \propto N. \label{R}
\eeq
Note that the {\it classical} SUGRA description is applicable when $R_3 \gg \sqrt{\alpha'}$,
that is when $g_{\rm s} N \gg 1$ (note that $g_{\rm s}\to 0$). Otherwise
string theory corrections are relevant and deform the soliton \eq{sol}.

  The geometry of the D3-brane soliton is as follows: It has asymptotically flat boundary
conditions at spacial infinity, as  $r \gg R_3$ the ratio $\left(R_3/r\right)^4$ gets much smaller
than unity. At the same time, near the position of the source ($r=0$) there is
an infinite throat region of a constant curvature:

\beqn
{\rm d}s^2 = \frac{r^2}{R_3^2} \, {\rm d}x_m \, {\rm d}x^m
+ \frac{R_3^2}{r^2} \, {\rm d}r^2 +
R_3^2 \, {\rm d}\Omega_5^2 \quad
{\rm with} \quad \exp{\left[- {\it \Phi}\right]} =
{\rm const}. \label{throat21}
\eeqn
By definition the throat is the region where $r \ll R_3$, so that
in \eq{fg} the unity can be neglected with respect to $\left(R_3/r\right)^4$.
In this way one obtains \eq{throat21} from \eq{sol}--\eq{fg}.

  As one can check directly, the metric \eq{throat21} has a constant scalar curvature
equal to $R_3$. The curvature does not diverge and
the D3-brane is a non-singular soliton. In fact, the metric \eq{throat21} has the geometry of
${\rm AdS_5\times S_5}$,
where ${\rm AdS_5}$ is five dimensional Anti-de-Sitter space and
${\rm S_5}$ is the five-sphere --- de-Sitter space.
Both of these manifolds are known to have constant scalar curvatures:
${\rm S_5}$ has positive while ${\rm AdS_5}$ has negative curvature. They are both solutions
to five-dimensional
Einstein equations with positive and negative cosmological constants, correspondingly.

   We now describe the geometry of ${\rm AdS_5}$ space.
There are many ways to present ${\rm AdS_5}$ space (see, for example, \cite{Witten}), but
we find the following description convenient. Algebraically ${\rm AdS_5}$ space could be represented as the {\it universal cover}
of a sub-manifold in a six-dimensional flat space ($W,V,X_q$ where $q=1,...,4$)
with signature ($-,-,+,+,+,+$).
The equation defining this sub-manifold is \cite{Witten}:

\beq
W^2 + V^2 - \sum_{q=1}^4 X_q \, X_q = R_3^2,\label{AdAd}
\eeq
where $R_3$ is the radius of the
sub-manifold in question and of ${\rm AdS_5}$
space. Thus, ${\rm AdS_5}$ admits a
natural action of the global SO(4,2), which is its isometry group. The metric on the ambient
flat six-dimensional space is:

\beq
{\rm d}s^2 = - {\rm d}W^2 - {\rm d}V^2 + \sum_{q=1}^4 {\rm d}X_q \,
{\rm d}X_q. \label{ammet}
\eeq
A metric on the universal cover of the manifold \eq{AdAd} can be found by
solving equation \eq{AdAd}:

\beqn
V &=& R_3 \, r \, t \nonumber \\
W &=& \frac{1}{2r}\left[1 + r^2 \left(R_3^2 + \sum_{q=1}^3 x_q^2 - t^2\right)\right] \nonumber \\
X_4 &=& \frac{1}{2r} \left[1 - r^2 \left(R_3^2 - \sum_{q=1}^3 x_q^2 + t^2\right)\right] \nonumber \\
X_q &=& R_3 \, r \, x_q, \,\, {\rm where} \,\, q=1,...,3 \label{AdCo}
\eeqn
Substituting this solution into equation \eq{ammet} we obtain the metric for
${\rm AdS_5}$ space:

\beq
{\rm d}s^2 = \frac{r^2}{R_3^2} \left(- {\rm d}t^2 + \sum_{q=1}^3 {\rm d}x_q \,
{\rm d}x_q\right) + \frac{R_3^2}{r^2} \, {\rm d}r^2,
\label{AdSMET}
\eeq
which coincides with the metric for the ${\rm AdS}_5$ part in \eq{throat21} if $x_m = (t,x_q)$ where $q=1,...,3$.

  Now let us define the boundary of ${\rm AdS_5}$ space. If $W,V,X_q$ (where $q=1,...,4$) go to infinity,
after dividing the coordinates by a positive constant one obtains an equation defining the
boundary:

\beq
W^2 + V^2 - \sum_{q=1}^4 X_q \, X_q = 0.\label{Mink}
\eeq
The boundary is a four-dimensional manifold, because \eq{Mink} is invariant under scalings
$W\to \lambda W, V \to \lambda V, X_q \to \lambda X_q$ for real non-zero $\lambda$.
Scaling by positive $\lambda$ one can map \eq{Mink} to the locus:

\beq
W^2 + V^2 = \sum_{q=1}^4 X_q \, X_q = 1, \label{Mink1}
\eeq
which is a copy of $\left(S^1\times S^3\right)/Z_2$. One must factor over $Z_2$ here
because there is a remaining symmetry under $W\to - W, V\to -V, X_q\to - X_q$.
Universal cover of \eq{AdAd} has as a boundary
the universal cover of \eq{Mink1}, which is $R^1 \times S^3$. The latter manifold is
a conformal compactification of the four dimensional Minkowski space $R^{3,1}$.
In fact, for the conformal compactification of $R^{3,1}$ one adds a point at spacelike infinity.
In terms of the metric \eq{AdSMET} this could be clarified as follows. There are two parts of the
${\rm AdS}_5$ boundary: first one is at $r\to \infty$, which is four-dimensional Minkowski space $(t,x_q)$
where $q=1,...,3$; second part of the boundary is a point $r = 0$.

   From these considerations it follows that there is a natural action
of SO(4,2) on the conformal compactification of the Minkowski space.
This group now defines four-dimensional conformal
transformations. Note that under a generic conformal transformation the point $r=0$
is mapped to a point inside $R^{3,1}$. That is the reason that the
compactification of $R^{3,1}$ is referred to as conformal.

 Note that SUGRA on ${\rm AdS}_5$ space is invariant under a global SO(4,2) symmetry.
Furthermore SUGRA on the throat \eq{throat21} of the D3-brane is invariant under
$\cN = 8$ SUSY.

  Now let us consider the SYM description of the D3-brane.
This description is applicable when $g_{\rm s}\to 0$, and
as follows from \eq{SYM}, the description is given by $\cN = 4$ four-dimensional
SYM:

\beqn
S = \frac{1}{4\pi g_{\rm s}} \int {\rm d}^4 x \, {\rm tr}
\left\{\frac12\hat{f}_{mn}^2 +
\frac12\left|D_m\hat{\phi}_i\right|^2 +
\frac12\sum^6_{i>j}\left[\hat{\phi}_i, \hat{\phi}_j\right]^2 +
\phantom{\frac12^{\frac12}}
\right. \nonumber \\ \left. \phantom{\frac12^{\frac12}} +
\frac{\rm i}{2}\sum^4_{I=1}\hat{\bar{\Psi}}^I\hat{D}\hat{\Psi}_I -
\frac{\rm i}{2}
\hat{\Psi}^I\left[\hat{\phi}_{IJ},\hat{\bar{\Psi}}^J\right] + {\rm c.c.}
\right\},\label{N=4}
\eeqn
where $\hat{\phi}^{IJ} = \hat{\phi}^i\gamma^{IJ}_i$
and $\gamma^{IJ}_i$ are six-dimensional Dirac matrices.  One can see from
this formula that $4 \, \pi \, g_{\rm s} = g^2$, and so when $g_{\rm s}\to 0$
the perturbative expansion of SYM is well defined. The theory \eq{N=4} has vanishing
$\beta$-function because of the perfect cancellation of quantum corrections
due to bosons and fermions. Hence, $g$ is just a non-renormalizable
constant, which is in accordance with the
fact that $g_{\rm s} = \exp{\left[2 \, {\it \Phi}\right]} = {\rm const}$.  Furthermore, at any value of
$g$ the theory is invariant under four-dimensional conformal
transformations given by SO(4,2) group.  The conformal symmetry extends
$\cN = 4$ SUSY invariance of SYM theory in question to $\cN = 8$ SUSY.

   This shows that SO(4,2) is naturally realized both on the SYM and SUGRA sides, which is
a good sign that $\cN = 4$ SYM theory should be related to type {\rm II}B
SUGRA on the ${\rm AdS_5\times S_5}$ space with self-dual
R--$\tilde{\rm R}$ four-form flux\footnote{The self-dual
R--$\tilde{\rm R}$ four-form flux is present becuase the ${\rm AdS_5\times S_5}$
geometry appears from the D3-brane which is charged with respect to this field.}
\cite{Maldacena}. Note that {\it classical} type {\rm II}B SUGRA description is valid when
$R_3/\sqrt{\alpha'} \to \infty$, which corresponds, according to \eq{R}, to
taking $N\to\infty$ as well as $g_{\rm s} N \to \infty$ (note that $g_{\rm s}\to 0$).
Hence, {\it strongly} coupled $\cN = 4$ SYM theory in the large $N$ limit
is applicable in absolutely the same situation as type {\rm II}B SUGRA on an
${\rm AdS_5\times S_5}$
background. These naive considerations favor a relation between
the two theories in question will receive a further support below.

\subsection{ABC of the AdS/CFT-correspondence}

   We now wish to present in a formal way the relation which
we are going to study below. The relation is between $\cN = 4$
four-dimensional ${\rm SU(N)}$ SYM and type {\rm II}B SUGRA in an
${\rm AdS_5\times S_5}$
background with R--$\tilde{\rm R}$ four-form  flux \cite{GuKlPo,Witten}.
It establishes that as $g_{\rm s} N \to \infty$, while
$g_{\rm s} \to 0$ and $N\to\infty$:

\beq
\left\langle \exp{\left[- {\rm i}\sum_j \int {\rm d}^4 x \, J_0^j(x) \, \cO^j
\right]} \right\rangle \approx
\exp\left\{- \, {\rm i} \, S^{\rm min}\left[\left({\rm AdS_5}\right)_N^{\phantom{\frac12}}
\times \,\, \left({\rm S_5}\right)_N\right]_{J^j|_u = J_0^j}\right\}. \label{boundary}
\eeq
The average on the LHS is taken in {\it strongly coupled} large $N$
SU$(N)$, $\cN = 4$  SYM theory; $\left\{\cO^j\right\}$ is a complete set
of local operators, which respects the symmetries of the problem.
On the RHS of \eq{boundary} $S^{\rm min}$ is type {\rm II}B SUGRA
action in an ${\rm AdS_5\times S_5}$ background with self-dual
R--$\tilde{\rm R}$ four-form flux. The action is
minimized on classical solutions for all its fields: note that as $R_3/\sqrt{\alpha'}
\propto g_{\rm s} N \to \infty$ string theory corrections to this SUGRA theory are suppressed. The classical
solutions in SUGRA are represented schematically as $J_j$: for example, $j$
could contain tensor indexes.  These solutions have values
$J^j|_u = J_0^j$ at the four-dimensional hyper-surface $r = u < R_3$ in the
${\rm AdS_5}$
space and some asymptotic behavior as $u\to R_3$ \cite{Witten}.  These values $J_0$ serve as sources in the LHS.
Thus, we see that type {\rm II}B SUGRA in the {\it bulk} of the ${\rm AdS_5}$ space is related to SYM
theory living on {\it four-dimensional hyper-surfaces} ($r=u$ for an arbitrary $u$) inside the space in question.
This is so called Holography phenomenon \cite{Susskind,tHoH,WiSu} in quantum field theory.

  Relations between the different parameters on both sides of \eq{boundary} are:

\begin{eqnarray}
R_3^4 &=& 4\pi g_{\rm s} N\alpha'^2 \nonumber\\
g^2 &=& 4\pi g_{\rm s} = {\rm const} \nonumber \\
M_{UV} &=& \frac{R_3}{\alpha'} \nonumber \\
{\rm Rank \,\, of \,\, the \,\, gauge \,\, group} &=& {\rm number \,\, of \,\, units
\,\, of \,\, R-\tilde{R} \,\, flux = N \propto Q_3}
\nonumber \\ \frac{u}{\alpha'} &=& {\rm energy \,\, scale \,\, on \,\, the \,\, SYM \,\, side},
\label{woka}
\end{eqnarray}
where $M_{UV}$ is an UV cutoff for SYM. In fact, the generating
functional of SYM correlation functions (LHS of \eq{boundary}) has UV
divergences and needs to be regularized. Hence, the SYM generating
functional evolves under renormalization group (RG) flow.
This is despite the fact that there are
no quantum corrections to the classical action \eq{N=4} of
four-dimensional $\cN = 4$ SYM theory. ${\rm AdS_5}$ SUGRA needs
to be regularized as well, as we discuss below, and natural regularization
parameter is again $R_3$ \cite{GuKlPo,Witten}.

  Before discussing the meaning of the relation \eq{boundary} let us emphasize
that it is similar to the relation \eq{26} between string theory and gravity.
In this case SUGRA theory appears as an effective theory
of SYM. One of the differences from the string theory statement \eq{26} is that now we get
``$Z = \exp{\left[- {\rm i} \, S({\rm sources})\right]}$'' because SYM is a {\it second quantized} theory.

  The relation between the two theories in question should be
understood as follows:  there is a quantum type {\rm II}B superstring theory on
${\rm AdS_5\times S_5}$
with a R--$\tilde{\rm R}$ background, which is valid at any energies and yet to
be found. This string theory is weakly coupled
when $g_{\rm s} \to 0$, so to keep $g_{\rm s} N$ fixed one should take $N\to\infty$.
At energies smaller than $R_3/\alpha'$
superstring theory in question has two degenerate limits,  one of which happens
when $g^2 N \propto g_{\rm s} N \ll 1$. It is described by weakly coupled $\cN = 4$ SYM at large $N$,
which is well defined theory. The other occurs when $R_3^4/\alpha'^2 \propto g_{\rm s} N \gg 1$. In this
limit one must deal with strongly coupled SYM theory whose definition is
not known. Then the proper description when $g_{\rm s} N \gg 1$
is given by weakly coupled ({\it classical}) type {\rm II}B SUGRA on the background in question.

\subsection{Interpretation}

   Consider now type {\rm II}B string theory in an ${\rm AdS_5 \times S_5}$
background with R--$\tilde{\rm R}$ flux
corresponding to the D3-brane. This theory is quantum gravity, therefore, in it one averages over all
metrics with the asymptotically AdS boundary conditions. As a result, correlation functions in this theory
are independent of the choice of metric. Hence, the correlators are independent of the coordinates
of the operators acting in the bulk of ${\rm AdS_5}$. Thus, all correlators in the theory for operators
placed in the bulk of ${\rm AdS_5}$ are trivial. Moreover, because
${\rm AdS_5}$ space does not contain an asymptotically flat part,
SUGRA in an ${\rm AdS}_5$ background is always strongly coupled
in the sense that there are no asymptotic states.

  Thus, it is natural to consider a quantity in ${\rm AdS_5}$ SUGRA which generates correlation
functions of operators acting at the boundary of ${\rm AdS_5}$ space.
This quantity is nothing but a wave-functional in the SUGRA theory.
The operators in question
should be those which create or annihilate various SUGRA particles at the boundary.
The classical limit of such a generating functional is the RHS of \eq{boundary}.
It is important that correlations between operators acting at the boundary of
${\rm AdS_5}$ are non-trivial. In fact, after fixing the boundary
in ${\rm AdS_5}$ space there is a natural \cite{Witten}
choice of metric on the boundary within the conformal class given by the bulk
metric\footnote{The boundary metric is obtained by multiplication of eq.
\eq{AdSMET} by $1/r^2$ and taking $r\to\infty$.} \eq{AdSMET}.

   In other words, gravity in ${\rm AdS_5}$ is entirely described by an
SO(4,2) ({\it conformaly}) invariant field theory
living only on its boundary, or on any four-dimensional hyper-surface with
$r=u \le R_3$. The generating functional we have considered
above for ${\rm AdS}_5$ gravity theory is equivalent to the generating
functional of a four-dimensional conformal field theory\footnote{Compare this
statement with \eq{boundary}.}. A question to be answered is what kind of conformal theory is living on the
four-dimensional hyper-surfaces in the ${\rm AdS_5}$ space?

   Now that we have established how the correspondence \eq{boundary} can be understood from
bulk theory point of view, let us clarify how the things are seen from the
boundary theory point of view. Defining the classical limit of the gravity generating functional at the boundary
one can find (via SUGRA equations of motion) its value at any hyper-surface $r=u$.
On the boundary theory side this is seen as a RG flow from the cutoff $R_3/\alpha'$ to the
energy scale $u/\alpha'$. In fact, the LHS of \eq{boundary} is nothing but the
Wilsonian effective action for the boundary theory, which is defined at the energy
scale $r=u$.  At the same time, the asymptotic behavior of sources (coefficient
functions) $J_0$ as $r=u \to R_3$ is given by
perturbative $\beta$-functions in the boundary theory. Note that coefficient functions
of the Wilsonian effective action depend not only on $u$ but also on the coordinates of
the four-dimensional space time ($x_m$). This fact is necessary for Holography to be valid
from the point of view of the theory confined to the boundary.

  To explain this consider that it is Holography which allows one to find the generating functional
in the boundary theory at the energy scale $u/\alpha'$ if one knows
the value of this functional at any other energy scale, independently of whether it is bigger
or smaller than $u/\alpha'$. For example, if one knows the generating functional
of the boundary theory at the energy scale $u/\alpha' < R_3/\alpha'$, then it is possible
to find its value at the cutoff scale $R_3/\alpha'$.

  Now we temporarily forget about the AdS/CFT-correspondence
and just look at what happens to the boundary theory.
In the RG evolution of this theory we integrate out high energy modes.
If in this integration one was only keeping information about divergent
counter-terms in the limit $R_3/\alpha'\to \infty$, there would
be no way to recover the UV theory from the IR one. In fact, there
could be many different UV theories which would flow
to the same IR one. This is in drastic contradiction with Holography.
To restore Holography one must keep all information about high energy modes in the RG evolution
of the theory. This is done by keeping {\it all} counter-terms,
even those which are finite as $R_3/\alpha'\to \infty$. In this way
all information about high energy modes is encoded in terms of {\it all}
sources $J_0$ provided the latter are only functions of $x_m$.
Specifically we mean that in the latter case any variation
of the fields in the theory could be compensated by a
variation of the sources $J_0(x)$. Thus, if one knows the values of all $J_0$ (i.e. one knows the
SYM generating functional) at some $r=u$ it is possible to
find them at any other $r=u_1$.

  Unfortunately, there is no rigorous derivation of the equality
\eq{boundary} and one can not straightforwardly trace the ``boundary'' theory.
Hence, the best that can be done now is to present
different points of view and to give some self-consistency arguments in
favor of the correspondence.

  Below we explain why SUGRA on the asymptotic flat space of the whole
D3-brane soliton should
decouple from the relation \eq{boundary}; why ${\rm AdS}_5$ SUGRA is related to
SU$(N)$ SYM rather than to U$(N)$;
why the limits $N\to\infty$ and $g_{\rm s} N\to \infty$ should be taken; why $R_3/\alpha'$ ($u/\alpha'$) plays the
role of the UV cut off (energy scale) in SYM theory; what specifies which field
in SUGRA is related to which operator in SYM and vise versa.

\subsection{Qualitative observations}

  Let us consider what is going on with the $N$ D3-brane bound state at very
low energies as measured by an observer at infinity
\cite{Maldacena,RMAdS}. According to \eq{bibi} in this limit the observer sees
free (non-interacting) ten-dimensional SUGRA in the bulk:
all interactions are suppressed, because $\Gamma_N$ is small with respect to
a characteristic scale in the theory. In fact:

\beq
S \propto \frac{1}{\Gamma_N} \int {\rm d}^{10} x \, \sqrt{-G} \, \cR + ... \propto
\int {\rm d}^{10}
x\, \left[\left(\diff h\right)^2 + \sqrt{\Gamma_N} \,
\left(\diff h\right)^2 \, h + ...
\right]. \label{free}
\eeq
Here we have parametrized the metric as $G = \eta +
\sqrt{\Gamma_N} h$, where $\eta$ is the flat metric and $h$ represents small
fluctuations around it.

  Because all interactions are suppressed, free SUGRA decouples
from the D3-brane excitations which are described by SU$(N)$ SYM \eq{N=4}.
Of all D3-brane excitations described by ${\rm U(N) = SU(N) \times U(1)}$
SYM those which correspond to the U$(1)$ part are not decoupled from
free SUGRA. In fact, they describe the center of mass
degrees of freedom and correspond to the source for the corresponding
D3-brane soliton. Hence, these excitations are coupled to bulk SUGRA even in the low energy limit.

  That is only one way of looking at the things. Another point of view is that
according to \eq{Dsur} and \eq{sol}, free SUGRA seen by the
observer at infinity is decoupled from the SUGRA living in the throat region \eq{throat21} of the
R--$\tilde{\rm R}$ D3-brane. In fact, the bulk massless particles decouple from the throat region,
because their low energy absorption cross section by the D3-branes goes like
\cite{Klebanov}:

\beq
\sigma \propto \omega^3 R_3^8, \label{crosig}
\eeq
where $\omega$ is the energy of an in-going scalar particle as measured by an observer at infinity.
The cross section vanishes as we lower $\omega$.
This behavior can be understood as follows: in the low energy limit the
wave-lengths of particles in the bulk become much bigger than the typical
gravitational size of the brane $R_3$. Hence, long wave-length fluctuations do not see
regions of size $\sim R_3$.

 At the same time, \eq{crosig} is equivalent to the grey-body factor for the
D3-brane soliton. In this language the behavior of the grey-body factor
\eq{crosig} could be understood as follows: As we lower the energy (as measured by a distant
observer) of the excitations whose wave-function is centered close to the
position of the brane ($r \ll R_3$), these excitations find it harder and harder to climb the gravitational
potential of the D3-brane and escape to the asymptotic region.
As a result, the throat region and asymptotic one do not interact
with each other in the low energy (as measured by a distant observer) limit.

   In conclusion, there are two pictures describing the same phenomenon.
In both cases we have two decoupled theories in the low energy limit from the point of view of a
distant observer. In both cases one of the decoupled theories is free SUGRA in the ten-dimensional flat space.
So, it is natural to identify the other two systems which appear in both descriptions \cite{Maldacena}.
The latter systems are $\cN = 4$ four-dimensional SU$(N)$ SYM and type
{\rm II}B SUGRA in an ${\rm AdS_5\times S_5}$ background with
self-dual R--$\tilde{\rm R}$ four-form flux.

   What is most important for the whole picture is that the two theories in question
possess finite ({\it non-zero}) energies \cite{Maldacena}.  In fact,
their energy scales are those which are seen by an observer in the throat (at a fixed $r$
less than $R_3$)
rather than those which are seen by an observer at infinity.
Note that the $g_{tt}$ component of the D3-brane
metric is not a constant.  Hence, the energy $E_r$ of an object as measured
at a constant position $r$ and the energy $E_{\infty}$ measured by an
observer at infinity are related by the red-shift factor:

\beq
E_{\infty} = f_3^{-1/4} E_r.
\eeq
From this it follows that the same object, having fixed finite energy, as brought
closer and closer to $r=0$ would appear to have smaller and smaller energy
to an observer at infinity.

\subsection{Additional arguments}

In this subsection we present a few more calculations which favor the
correspondence \eq{boundary}.

{\bf 1.} First, we explain how one finds relations between operators on the
LHS and fields on the RHS of \eq{boundary}. For the massless excitations in
SUGRA one could use \eq{BI} or its non-Abelian generalization \cite{Tseytlin}.
Take for example the dilaton field. It couples to SYM as follows:

\beq
\Delta_{{\it \Phi}} S \propto \int {\rm d}^4 x \,
\exp{\left[- {\it \Phi}(x_m,\phi_i)\right]} \,
\left[f_{mn}^2 + \sum_{i=1}^6|\diff_m \phi_i|^2 + {\rm fermions} \right]. \label{dilfil}
\eeq
Note that the dilaton field depends on the $\phi_i$ fields in addition to $x_m$. That is to say the dilaton field is a function
of all ten coordinates rather than only of four $x_m$. We are going to consider small fluctuations of the dilaton
field around the background \eq{throat21}. Hence, we expand $\exp{\left[- {\it \Phi} \right]}$ in powers of the dilaton field
and the field itself in powers of $\phi_i$. Then from \eq{dilfil} we obtain:

\beq
\Delta_{{\it \Phi}} S_n \propto \int {\rm d}^4 x \, \diff_{i_1}...\diff_{i_n}
{\it \Phi}(\phi_i, x)|_{\phi_i = 0}
\,\left( \phi_{i_1} ... \phi_{i_n}
\, \left[f_{mn}^2 + \sum_{i=1}^6|\diff_m \phi_i|^2 + {\rm fermions}\right]\right).
\eeq
We can see from this that the $n$-th spherical harmonic of the dilaton field in
${\rm S_5}$ (i.e. a KK mode in ${\rm S_5}$)
couples to the operator

$$
\cO_n^{\it \Phi}\left[\phi_i, a_m\right] \propto
\phi_{i_1} ... \phi_{i_n} \, \left[f_{mn}^2 + \sum_{i=1}^6|\diff_m \phi_i|^2
+ {\rm fermions}\right]
$$
The non-Abelian generalization of this operator is:

\beq
\cO_n^{\it \Phi}\left[\hat{\phi}_i, \hat{a}_m\right] \propto {\rm tr}\left(\hat{\phi}_{(i_1} ... \hat{\phi}_{i_n)}
\, \left[\hat{f}_{mn}^2 + \sum_{i=1}^6|D_m \hat{\phi}_i|^2 + \frac12\sum_{i>j} \left[\hat{\phi}_i,
\hat{\phi}_j\right]^2 + {\rm fermions}\right]\right).
\eeq
One can conclude from this that the zero mode of the dilaton field ($n=0$)
couples to the SYM action \eq{N=4}.
Similarly from \eq{BI} one can find that the zero
mode of the graviton field $G_{mn}(x,\phi_i = 0)$ couples
to the SYM energy-momentum tensor.

  In general the method of finding relations between SUGRA fields and
SYM operators is based on matching of their symmetry properties under the group
SO(4,2) \cite{RMAdS}.  Remarkably, it appears that
for each SUGRA field in the chiral representation of (the SUSY extension of)
SO(4,2) group there is a SYM operator transforming in the same
representation \cite{RMAdS} and vice versa.

  It is worth mentioning at this point that there are other
symmetry arguments in favor of the validity of the AdS/CFT-correspondence
\cite{Ferrara,RMAdS}, though, we are not going to discuss them here.

{\bf 2.}  Second, bearing the above considerations in mind, let us examine the relation
\eq{boundary} in more details. Following \cite{GuKlPo,Witten},  we consider
the zero mode of the dilaton. The action for a dilaton field in the
${\rm AdS_5}$ background
in the linear approximation is \cite{GuKlPo,Witten}:

\beq
S({\it \Phi}) = \frac{\pi^2 R^8_3}{32 \Gamma_N} \, \int {\rm d}^4 x \, d z \, \frac{1}{z^3} \, \left[
\left(\diff_z{\it \Phi}\right)^2 + \left(\diff_m {\it \Phi}\right)^2
\right] + ... \label{clac}
\eeq

Here the metric on the ${\rm AdS_5}$ space is taken as:

\beq
{\rm d}s^2 = \frac{R_3^2}{z^2} \left({\rm d}z^2 + \eta^{mn} \, {\rm d}x_m \,
{\rm d}x_n\right), \quad {\rm where} \quad
z = \frac{R_3^2}{r}. \label{AdSMET1}
\eeq
In this metric the boundary of ${\rm AdS}_5$ space is Minkowski space at $z=0$ plus a point at $z\to\infty$.

The action \eq{clac} is divergent for those classical solutions which are regular on the
boundary and fall off for large $z$ \cite{GuKlPo,Witten}. To regularize this divergence
there is a natural cutoff of ${\rm AdS}_5$ space
at $z= \epsilon \propto \alpha'/R_3$. This is an infrared (IR) regularization of
AdS SUGRA.

  Now any classical solution with ${\it \Phi}(z= \epsilon,x) =
{\it \Phi}_0(x)$
could be expanded in those which obey ${\it \Phi}(z=\epsilon,x) = \exp{\left[{\rm i} \,  k_m \, x^m\right]}$,
where $k_m$ is a four-momentum \cite{Witten}.  The unique normalizable \cite{GuKlPo,Witten}
solution with the latter boundary condition and which is regular
as $z\to\infty$ is \cite{GuKlPo,Witten}:

\beq
{\it \Phi}(x_m, z) = \frac{(kz)^2 \, \cK_2(kz)}{(k \epsilon)^2 \, \cK_2(k \epsilon)}
\, \exp{\left[{\rm i} \, k_m \, x^m\right]} \quad {\rm where} \quad k = |k_m|.
\eeq
Here $\cK_2$ is the modified Bessel function.  The action for this solution
is equal to \cite{GuKlPo,Witten}:

\beqn
S^{\rm min}({\it \Phi}_0) \propto N^2 \int {\rm d}^4 x \, \int {\rm d}^4 y \,
{\it \Phi}_0(x) \, {\it \Phi}_0(y) \, \frac{1}{\left(\epsilon^2 + |x_m -
y_m|^2\right)^4} + O(\epsilon^2),  \label{regac}
\eeqn
where ${\it \Phi}_0(x) = \exp{\left[{\rm i} \, k_m \, x^m \right]}$. We have got a prefactor of $N^2$
in the integrals \eq{regac} because $R_3^2 \propto N^2$.
There is no contibution to \eq{regac}, which is of the order of ${\it
\Phi}_0^2$,
from higher order corrections in \eq{clac}.

  At the same time, the generating functional in the SYM picture is:

\beqn
Z({\it \Phi}_0) = \int \cD \hat{a}_{m} \, ... \, \exp\left\{-
\frac{\rm i}{g^2}\int {\rm d}^4 x\, {\rm tr} \left[\hat{f}_{mn}^2 + ...\right] \, +
\, \frac{\rm i}{g^2} \int {\rm d}^4 x \, {\it \Phi}_0(x) \, {\rm tr}
\left[\hat{f}_{mn}^2 + ...  \right]\right\}. \label{genSYM}
\eeqn
Dots in this equation stand for the superpartners of $a_m$.
Now ${\it \Phi}_0(x)$ is a source for the operator which is the
SYM classical action. According to \eq{boundary} and
\eq{BI} it should be equal to the dilaton's boundary value
${\it \Phi}(z= \epsilon,x)$.

  Integrating over the SYM fields in \eq{genSYM}, we get:

\beqn
Z({\it \Phi}_0) = {\rm const} \cdot \exp \Bigl\{- {\rm const} \cdot  {\rm i} \,
\int {\rm d}^4 x \int {\rm d}^4 y~~{\it \Phi}_0(x) \cdot
{\it \Phi}_0(y) \times \nonumber \\ \times \left\langle {\rm tr} \left[\hat{f}_{lm}^2(x) +
... \right]\, {\rm tr}\left[\hat{f}_{np}^2(y) + ...\right]\right\rangle + ... \Bigr\}.
\label{gen}
\eeqn
up to {\it quadratic} order in the dilaton. Because of
restrictions imposed by $\cN = 4$ SUSY invariance we know the
exact value of the correlator:

\beqn
\left\langle{\rm tr}\left[\hat{f}_{lm}^2(x) + ...\right]\, {\rm tr}\left[\hat{f}_{np}^2(y) +
...\right]\right\rangle \propto \frac{N^2}{\left|x_m - y_m\right|^8}.
\label{regsc}
\eeqn
Here $N^2$ appears as the number of degrees of freedom in SYM theory. In fact,
$\cN = 4$ SYM theory is superconformal and, hence is not confining:
the degrees of freedom are the same at {\it all} scales.

  Now in eq. \eq{gen}, \eq{regsc} there is an UV divergence when
$x=y$.  It can be regularized via point splitting. Concisely this means
that all distances in four-dimensional space-time must be bigger than
some regularization parameter $\epsilon'$.  In this regularization scheme we
have:

\beqn
\left\langle{\rm tr}\left[\hat{f}_{lm}^2(x) + ...\right]\, {\rm tr} \left[\hat{f}_{np}^2(y) +
... \right]\right\rangle \propto \frac{N^2}{\left(\epsilon'^2 + |x_m -
y_m|^2\right)^4} + {\rm contact \,\, terms}.
\eeqn

   In conclusion, if we equate $\epsilon' = \epsilon$, we find an
agreement between the LHS and RHS of \eq{boundary}.
Furthermore, we find that the IR regularization on the SUGRA
side is related to the UV one in SYM \cite{GuKlPo,Witten}.
Thus, $R_3/\alpha'$ plays the role of a UV regularization
on the SYM side. Note that one could vary the SYM UV regularization parameter as well
as the position of the boundary of the ${\rm AdS}_5$ space by SO(4,2) transformations.
In other words, one could place the hyper-surface on which SYM
lives to any position $r=u$ inside ${\rm AdS}_5$ space by an SO(4,2) transformation.

The check we just performed could also be extended to other SYM
operators and SUGRA fields to obtain other agreement \cite{RMAdS}.

{\bf 3.}  Third, at this point one could ask: What is
the meaning of the string theory for $\cN = 4$ SYM? Normally such a
string representation means confinement in the theory \cite{Polyakov}.
In fact, consider the Wilson loop:

\beq
W(C) = {\rm \, tr \, P} \exp{\left[{\rm i} \, \oint_C d x_m \, \hat{a}_m\right]}.
\eeq
In this formula $C$ is some contour inside four-dimensional space-time
and the trace is taken in the fundamental representation of the
gauge group.

   The string representation of YM theory means that the Wilson loop
expectation value could be represented as a sum over string world-sheets
$\Sigma_C$ having $C$ as their boundary:

\beq
\langle W(C) \rangle = \sum_{\Sigma_C} \exp{\left[- {\rm i} \, S(\Sigma_C)\right]}
\label{WIL}
\eeq
for some string theory action $S(\Sigma_C)$. Regularly this
tells us that in Euclidean space if one takes a large loop $C$ then this
becomes:

\beq
\langle W(C) \rangle \propto \exp{\left[- \cA(\Sigma_C^{\rm min})\right]}, \label{arealaw}
\eeq
where $\cA(\Sigma_C^{\rm min})$ is an area of the minimal surface $\Sigma_C^{\rm min}$
spanned by $C$. This signals linear potential between the sources in the
fundamental representation of the gauge group and hence confinement
\cite{Polyakov}.

  One can use the AdS/CFT-correspondence in Euclidean space
\cite{Kor,Malwil,WitT} to find a
representation like \eq{arealaw} for the Wilson loop expectation value
in $\cN = 4$ SYM. The answer for strongly coupled
SYM theory is the same as in \eq{arealaw}, but now
$\cA(\Sigma_C^{\rm min})$ is a regularized \cite{Kor,Malwil} area of the minimal surface
spanned by the contour $C$. The latter now lives on the boundary of AdS
space. At the same time, the string world-sheet lives
inside ${\rm AdS}_5$ space.

   Note that we do not expect confinement for a conformal theory,
because one has the same degrees of freedom in such a theory
at all scales. Thus, the question appears:
why does an answer like \eq{arealaw} for the Wilson loop average
in $\cN = 4$ SYM theory not lead to confinement?
In other words, as the area enclosed by $C$ on the boundary is scaled up,
why is the area $\cA(\Sigma_C^{\rm min})$ not scaled
up proportionately? It is the AdS geometry that is helpful \cite{WitT}.
In fact, the answer to this question is clear from SO(4,2) invariance:
If we rescale $C$ by $x_m\to t x_m$, with a large positive $t$,
then by conformal invariance we can rescale $\Sigma_C^{\rm min}$, by $x_m \to t x_m$
and $z\to t z$ (see \eq{AdSMET1}), without changing its area $\cA$. Thus the area $\cA$
need not be proportional to the area enclosed by
$C$ on the boundary. Since, however, in this process we had to scale $z\to t z$
with very large $t$, the surface $\Sigma_C^{\rm min}$ which is bounded by
a very large circle $C$ should extend very far away from the boundary
of ${\rm AdS}_5$ space. This is perfectly consistent with AdS geometry.
Direct calculation in \cite{Kor,Malwil} shows that these
considerations are correct.

  There exist other arguments in favor of the validity of the
AdS/CFT-correspondence \cite{RMAdS}, but we stop here, since we hope that this is
enough to convince the reader that the AdS/CFT-correspondence should be
correct.

\section{Conclusions and Acknowledgments}

   Thus we see that the AdS/CFT-correspondence gives a first example of
a string theory description of SYM.  It is worth mentioning that analogues of the
AdS/CFT-correspondence could be established also for
SYM theories in other dimensions \cite{MaYa}. Moreover, it could be
generalized to conformal YM theories with less SUSY
\cite{KaSi,VaNe}. There are generalizations of the
AdS/CFT-correspondence for non-conformal theories \cite{WitT,KlTs,PolO}.

   Furthermore as is usual for such kinds of statements, which
relate two seemingly unrelated theories, this correspondence is useful for
both of its constituents \cite{RMAdS}. Besides the fact that the correspondence
suggests a string description of SYM, it gives a quantum description of gravity
in terms of SYM.  We mean that at distances much smaller than the string characteristic one (when
$g^2 N \ll 1$) we have a SYM description of quantum gravity: as we
mentioned the AdS SUGRA appears as an effective theory for SYM. Also, as
we noticed above, the AdS/CFT-correspondence gives an explicit example of the
Holography phenomenon, which can be important for understanding of
quantum gravity.

  For integrity we would like to criticize the status of the whole subject.
First, we see that it is possible to find a string description of YM theory
only in the most simplified situation.  In fact, the string description is
found when YM theory is maximally supersymmetric, when the large $N$ limit is
taken and it is more or less testable only for the strong coupling $g^2
N \to \infty$.  Second, even in the latter situation the correspondence is
not derived rigorously from first principles.

  My understanding of the D-brane physics and of the AdS/CFT-correspondence
was formed during discussions with Anton Gerasimov. I would like
to thank him for sharpening of my understanding of the subject. Also I am
indebted to N.Hambli, M.Laidlow, A.Losev, J.Maldacena, A.Marshakov,
A.Morozov, T.Pilling, R.Scipioni, G.Semenoff and K.Zarembo for
useful comments and discussions. This work was done under the support of
NSERC NATO fellowship grant and under the partial support of grants
INTAS-97-01-03  and RFBR 98-02-16575.

\section{Appendix. BPS states}

  In this appendix we define for completeness BPS solitons. We
present here a standard simple exercise \cite{WiOl} which could, as we hope,
help in the understanding why BPS solitons so special.

  Let us consider two-dimensional scalar SUSY theory:

\beq
S = \int {\rm d}^2\sigma \left[\frac12\left(\diff_a\phi\right)^2 + \frac{\rm i}{2}
\, \bar{\Psi} \, \hat{\diff}\Psi - \frac12 \, V^2(\phi) -
\frac12 \, V'(\phi)\, \bar{\Psi} \, \Psi\right],
\eeq
where $\Psi$ is a Majorana
fermion, and $V(\phi)$ is an arbitrary function (it could be $V =
-\lambda\left(\phi^2 - \phi_0^2\right)$ or $V(\phi) = - \sin\phi$, for
example).  The theory is invariant under SUSY transformations with conserved
Neuther current:

\beq
s^a = \left(\diff_b\phi\right) \, \gamma^b \, \gamma^a \, \Psi + {\rm i} \,
V(\phi) \, \gamma^a \, \Psi.
\eeq
Working with the chiral components $\Psi^{\pm}$ of the Fermi field, the chiral
components $Q^{\pm}$ of the SUSY charge can be written as follows:

\beqn
Q_{\pm} = \int {\rm d}\sigma_2 \, \left| \phantom{\frac12^{\frac12}} \left(\diff_1\phi \, \,
\pm \, \, \diff_2\phi\right)\Psi_{\pm} \mp V(\phi) \Psi_{\mp} \phantom{\frac12^{\frac12}} \right|.
\eeqn
In this notation the SUSY algebra is:

\beqn
Q_+^2 = p_+, \quad Q_-^2 = p_-, \quad Q_+Q_- + Q_-Q_+ =
\int {\rm d}\sigma_2 \, 2 \, V(\phi) \, \frac{\diff\phi}{\diff\sigma_2} \nonumber \\
{\rm where} \quad p_{\pm} = p_1 \pm p_2. \label{alg}
\eeqn
The RHS of the third equality here is so called central charge $\cZ$ of
the SUSY algebra.  It is proportional to the topological charge in the
theory. In fact, for example, if $V(\phi) = - \sin\phi$, then

$$
\cZ = \int^{+\infty}_{-\infty} {\rm d}\sigma_2 \, \frac{\diff}{\diff\sigma_2} \, \left(2
\cos\phi\right).
$$
The latter is non-zero only for (anti-) kink solutions. From
the algebra \eq{alg} one could find that:

\beq
p_+ +  p_- = \cZ + \left(Q_+ - Q_-\right)^2 = -
\cZ + \left(Q_+ + Q_-\right)^2,  \label{bps}
\eeq
hence, $p_+ + p_- \geq |\cZ|$. For a single particle state with mass $M$ at
rest this implies

\beq
p_-= p_+ = M \geq \frac12 |\cZ|. \label{bps1}
\eeq
This bound is saturated for the BPS states, when as seen from \eq{bps}
$$
\left(Q_+ + Q_-\right)|{\rm BPS}\rangle = 0, \quad
{\rm or} \quad
\left(Q_+ - Q_-\right)|{\rm BPS}\rangle = 0.
$$
For example, this condition is satisfied for all kink and anti-kink solutions of
this theory. Thus, the BPS states compose small representations of the SUSY
algebra: some combination of supercharges act trivially on the state, and
hence does not generate superpartners \cite{WeBa}.

  The last feature of the BPS states is crucial. In fact, if SUSY is not
broken (which could be checked from the beginning by calculation of the
Witten index for the theory), adiabatic variations of the parameters of the
theory do not change representations of the SUSY algebra. Hence, if the
equality \eq{bps1} holds at some values of the parameters, then it holds
always and BPS states survive quantum corrections.  Moreover one could
control the renormalization of the mass and charge using the equality
\eq{bps1}, and if there is enough SUSY neither mass nor charge are renormalized
at all.

\end{document}